\documentstyle[prb,aps,twocolumn,floats,epsf]{revtex}

\newcommand{\nn}{\nonumber \\}
\newcommand{\etal}{\textit{et al.}}

\begin{document}

\wideabs{

  \title{Multiparticle ring exchange in the Wigner glass and its
    possible relevance to strongly interacting two-dimensional electron
    systems in the presence of disorder}

  \author{Klaus Voelker}
  \address{Theoretische Physik, ETH H\"onggerberg,
           CH-8093 Z\"urich, Switzerland}
  \author{Sudip Chakravarty}
  \address{Department of Physics and Astronomy, University of
           California, Los Angeles, CA 90095}

  \date{\today}

  \maketitle

  \begin{abstract}

    We consider a two-dimensional electron or hole system at zero
    temperature and low carrier densities, where the long-range
    Coulomb interactions dominate over the kinetic energy. In this
    limit the clean system will form a Wigner crystal. Non-trivial
    quantum mechanical corrections to the classical ground state lead
    to multiparticle exchange processes that can be expressed as an
    effective spin Hamiltonian involving competing interactions.
    Disorder will destroy the Wigner crystal on large length scales,
    and the resulting state is called a Wigner glass. The notion of
    multiparticle exchange processes is still applicable in the Wigner
    glass, but the exchange frequencies now follow a random
    distribution. We compute the exchange frequencies for a large
    number of relevant exchange processes in the Wigner crystal, and
    the frequency distributions for some important processes in the
    Wigner glass. The resulting effective low energy spin Hamiltonian
    should be the starting point of an analysis of the possible ground
    state phases and quantum phase transitions between them.  We find
    that disorder plays a crucial role and speculate on a possible
    zero temperature phase diagram.

  \end{abstract}

  \pacs{}

  }

\section{Introduction}

In recent years two-dimensional electron or hole systems with very low
densities are intensively studied.\cite{RMP} Such systems can be generated at
the interface of gallium arsenide heterostructures or silicon
metal-oxide-semiconductor field effect transistors, and more recently
also in organic C$_{60}$ and polyacene films.\cite{Batlogg} These
materials provide an excellent environment to study the effects of
strong electron-electron interactions and disorder. One example is the
unexpected metal-insulator transition.\cite{RMP}

We consider two-dimensional electron or hole systems at zero
temperature and zero magnetic field. In the absence of disorder, it is
known that the system will form a Wigner crystal in the limit of very
low densities, where the non-trivial correlations can be described in
terms of multiparticle exchange processes.\cite{Herring,Thouless} The
exchange frequencies then determine the magnetic Hamiltonian. A
calculation of the exchange frequencies of a pure two-dimensional Wigner
crystal was pioneered by Roger.\cite{Roger}

Although conceptually important, the pure Wigner crystal cannot be
realized in the systems mentioned above, due to disorder.\cite{Tsui} A
measure of disorder is the Drude conductance at an
intermediate temperature scale at which the resistivity is relatively
flat as a function of temperature, and the dominant contribution is from
impurity scattering. At low densities, the measured Drude
conductances are of order $e^2/\hbar$, indicating the importance of
disorder. We consider this
intermediate-temperature conductance as a tuning parameter for the
quantum phase transitions to be discussed, not the asymptotic low
temperature conductance.  This characterization of the tuning parameter
is important because, even for a pure system, the conductance at a 2D
quantum critical point can be of order $e^2/\hbar$.\cite{Cha}

It is also known that even an arbitrarily small amount of disorder
will destroy the long-range order of the Wigner lattice.\cite{Larkin} On
short length scales, however, the lattice will remain unaffected by weak
disorder, so that the notion of the multiparticle exchange is still
valid. Strong disorder will compromise the crystalline order even on
length scales comparable to the lattice spacing. Nonetheless, the
multiparticle exchange picture depends only on the existence of a rigid
ground state in the classical limit (that is in the low density limit),
which can be assumed to hold for any disorder strength. The exchange
frequencies will, of course, follow a random distribution in the
presence of disorder.

In a previous paper \cite{PhilMag} on the metal-insulator transition,
we calculated a set of relevant exchange frequencies for the clean
Wigner crystal within the many dimensional WKB
approximation.\cite{Schmid} This allowed us to conjecture a possible
phase diagram in the ground state. The purpose of the present paper is
to extend this calculation to the random distribution of exchange
frequencies, necessarily caused by disorder in realistic situations.
The resulting random and competing magnetic Hamiltonian should be an
important ingredient in determining the phase diagram of this
correlated complex system. A recent numerical calculation of exchange 
constants in a clean Wigner crystal is also available.\cite{Ceperley} 
\subsection{Wigner crystal and Wigner glass}
\label{sec: WignerCystal}

\newcommand{\meff}{m^{*}} \newcommand{\EE}{E_{coul}}

A two-dimensional electron system with carrier density \( n_{s} \) is
characterized by the dimensionless parameter
\begin{equation}
\label{eq_rsab}
  r_{s}^{-1}=a_{B}(\pi n_{s})^{1/2},
\end{equation}
which is a measure of quantum fluctuations; larger $r_s$ implies
smaller quantum fluctuations. Here \( a_{B}=\hbar^2 \epsilon /\meff
e^2 \) is the effective Bohr radius; $m^*$ is the effective mass, and
$\epsilon$ is the background dielectric constant. Thus, $r_s$ is the
mean spacing between the carriers, in units of the Bohr radius. In a
dilute system, where \( r_s \) is large, we expect the ground state to
be determined by the electrostatic repulsion between the electrons. In
the absence of disorder, the classical ground state that minimizes the
potential energy is a triangular lattice, the Wigner
crystal.

The crystalline state can be approximately described in terms of
single-particle wavefunctions that locally resemble harmonic
oscillator wavefunctions. The spatial extent of these wavefunctions, $\Delta r$,
depends on the oscillator frequency as \( \Delta r \sim \omega
_{0}^{-1/2} \), where \( \omega_{0}^{2} \) is determined by the second
spatial derivative of the electrostatic potential. A dimensional
analysis yields \( \omega_{0}\sim r_{s}^{-3/2} \), so that \( \Delta
r/r_{s} \sim r_{s}^{-1/4} \), and the system becomes increasingly
classical as \( r_{s}\to \infty \) \( (n_{s}\to 0) \). At low
densities, we can therefore systematically expand around the classical
limit.

As the density increases, or $r_s$ decreases, the Wigner crystal will
melt at zero temperature. The melting transition in $(d+1)$-dimension,
where $d>1$, is likely to be discontinuous from Landau theory formulated in terms of the ground state
energy, which must be a unique functional, $E[\rho({\bf r})]$, of the density, $\rho({\bf
r})$, of the electron gas.\cite{Kohn} For a crystalline state, we can write
\begin{equation}
\langle \rho({\bf r})\rangle =\rho_0+\sum_{{\bf G}\ne 0}\rho_{\bf G}e^{i{\bf G}\cdot{\bf r}},
\end{equation}
where $\rho_0$ is the average density and ${\bf G}$'s are the
reciprocal lattice vectors of the crystal. In mean field theory, we
can consider the ground state energy to be simply a function of the
order parameters $\rho_{\bf G}$. Thus, the energy can be expanded as
\begin{eqnarray}
E&=&E[\rho_0]+\frac{1}{2}\sum_{\bf G}a_{\bf G}|\rho_{\bf G}|^2\nonumber \\
&+&u_3\sum_{{\bf G}_1,{\bf G}_2,{\bf G}_3}\rho_{{\bf G}_1}\rho_{{\bf G}_2}\rho_{{\bf G}_3}\delta_{{\bf
G}_1+{\bf G}_2+{\bf G}_3,0}\nonumber\\   &+&u_4\sum_{{\bf G}_1,{\bf G}_2,{\bf G}_3,{\bf
G}_4}\rho_{{\bf G}_1}\rho_{{\bf G}_2}\rho_{{\bf G}_3}\rho_{{\bf G}_4}\delta_{{\bf G}_1+{\bf G}_2+{\bf
G}_3+{\bf G}_4,0}\nonumber \\
&+&\cdots
\end{eqnarray}
The quadratic term is chosen to be
\begin{equation}
a_{\bf G}=a(r_s^c-r_s)+a'(G^2-k_0^2)^2,
\end{equation}
where $a$ and $a'$ are positive constants and $k_0$ fixes the length
of the reciprocal lattice vectors of the crystal. For simplicity,
$u_3$ and $u_4$ were chosen to be momentum independent, but functions
of $r_s$. On a triangular lattice, the cubic term is allowed by
symmetry, hence the transition to the crystalline state is
discontinuous in the order parameter, or ``first order". 

Consider now the regime of the phase diagram for $r_s>r_s^c$ and weak
disorder. We can prove that no matter how weak the disorder is the
crystal falls apart at the macroscopic scale. It is sufficient to
consider the limit $r_s\gg 1$, because quantum fluctuations can only
destabilize the crystal further.  We can now apply the famous
Imry-Ma-Larkin\cite{Larkin} argument. The gain in the pinning energy
due to disorder is proportional to $L^{d/2}$, whereas the cost in the
elastic energy of the crystal is $L^{d-2}$, where $L$ is the linear
dimension of the sample and $d$ is the space dimensionality. Thus, for
$d<4$, the pinning energy wins, and the crystal is destroyed for
arbitrarily small disorder. Even if the crystal is disordered in the
conventional sense, it still leaves open the possibility of a
power-law ordered state,\cite{Giamarchi1} but this is now proven not
to be possible in $d=2$.\cite{DSFisher} The density-density
correlation function falls off exponentially with a correlation
length, $\xi_D$, given by\cite{Giamarchi2}
\begin{equation}
\xi_D > R_a \exp[c\sqrt{\ln(R_a/a)}].
\end{equation}
where $R_a$ is the length at which the displacement of the lattice
becomes of the order of the lattice spacing $a$.  Precise calculations
of the positive constant $c$, $R_a$, or the prefactor are not known
for the Wigner crystal. Nonetheless, $\xi_D$ is likely to be a large
length in the limit of weak disorder, and it is safe to assume that
short-range crystalline correlations will survive.

In $d=2$, the lack of crystalline order, or even a power-law
crystalline order, in the presence of disorder, does not allow us to
argue for a distinct state of matter distinguished by its special
features with respect to the translational degrees of freedom. From
this perspective, one can continuously connect the liquid state and
the amorphous crystalline state by moving into the disorder plane.  Thus, in $d=2$,
the global symmetries that can be truly broken in the presence of disorder
are the spin rotational invariance, $\cal S$, the time reversal
invariance $\cal T$, and the gauge invariance $U(1)$. These symmetries
can still label many distinct states of matter. For a related perspective
on the problem of a pinned Wigner crystal in a magnetic field, see
Ref.~\onlinecite{Chitra}. We
note that in $d=3$  a power-law
ordered Wigner glass can exist as a distinct state of matter. 

\subsection{Magnetism: pure system}
\label{sec: magpure}
In discussing the magnetism of the insulating Wigner crystal, we shall
ignore anharmonicities of the zero point phonon degrees of freedom,
which may merely renormalize the exchange constants.  The low lying
magnetic Hamiltonian is due to tunneling of electrons between the
lattice sites and can be expressed in terms of the $p$-particle cyclic
permutaion operators $P_{1\ldots p}$. Thus,
\begin{eqnarray}
\label{eq_Hpics}
H&=&J_2\sum_{\begin{picture}(17,10)(-2,-2)
        \put (0,0) {\line (1,0) {12}}
        \put (0,0) {\circle*{5}}
        \put (12,0) {\circle*{5}}
\end{picture}}
 \left(P_{1,2}+P^{-1}\right)-J_3
\sum_{\begin{picture}(26,15)(-2,-2)
        \put (0,0) {\line (1,0) {12}}
        \put (0,0) {\line (3,5) {6}}
               \put (12,0) {\line (-3,5) {6}}
        \put (6,10) {\circle*{5}}
        \put (0,0) {\circle*{5}}
        \put (12,0) {\circle*{5}}
\end{picture}}
\left(P_{1,2,3}+P^{-1}\right)\nonumber\\
&+&J_4
\sum_{\begin{picture}(26,15)(-2,-2)
        \put (0,0) {\line (1,0) {12}}
        \put (6,10) {\line (1,0) {12}}
        \put (0,0) {\line (3,5) {6}}
        \put (12,0) {\line (3,5) {6}}
        \put (6,10) {\circle*{5}}
        \put (18,10) {\circle*{5}}
        \put (0,0) {\circle*{5}}
        \put (12,0) {\circle*{5}}
\end{picture}}
\left(P_{1\ldots
4}+P^{-1}\right)-J_5
\sum_{\begin{picture}(26,15)(-2,-2)
        \put (0,0) {\line (1,0) {24}}
        \put (6,10) {\line (1,0) {12}}
        \put (0,0) {\line (3,5) {6}}
        \put (18,10) {\line (3,-5) {6}}
        \put (6,10) {\circle*{5}}
        \put (18,10) {\circle*{5}}
        \put (0,0) {\circle*{5}}
        \put (12,0) {\circle*{5}}
        \put (24,0) {\circle*{5}}
\end{picture}}
\left(P_{1\ldots5}+P^{-1}\right)\nonumber \\ &+&J_6
\sum_{\begin{picture}(26,30)(-2,-15)
        \put (6,10) {\line (1,0) {12}}
        \put (6,-10) {\line (1,0) {12}}
        \put (0,0) {\line (3,5) {6}}
        \put (0,0) {\line (3,-5) {6}}
        \put (18,10) {\line (3,-5) {6}}
        \put (18,-10) {\line (3,5) {6}}
        \put (6,10) {\circle*{5}}
        \put (6,-10) {\circle*{5}}
        \put (18,10) {\circle*{5}}
        \put (18,-10) {\circle*{5}}
        \put (0,0) {\circle*{5}}
        \put (12,0) {\circle*{2}}
        \put (24,0) {\circle*{5}}
\end{picture}
}\left(P_{1\ldots
6}+P^{-1}\right)+\cdots.
\end{eqnarray}
The sums are over the permutations shown in this equation.  There is a
theorem due to Herring and Thouless that exchanges involving even number of
fermions are antiferromagnetic, and those involving odd number of
particles are ferromagnetic.\cite{Thouless} We shall follow the
convention that the $J$'s are all positive.

A tractable method for calculating the exchange constants $J_2$,
$J_3$, \ldots\ is the instanton (or the many dimensional WKB)
method. It will be shown that $J_p$ is
\begin{equation}
  \label{WKB}
  J_p=A_p\hbar\omega_0 \left( \frac{S_p}{2\pi\hbar}
  \right)^{1/2} e^{-S_p/\hbar},
\end{equation}
where $S_p$ is the value of the Euclidean action along the minimal
action path that exchanges $p$ electrons. The quantity $\omega_0$ is
the characteristic attempt frequency, which can be estimated from the
phonon spectrum of the lattice. The prefactor $A_p$ is of order unity,
and the Eq.~(\ref{WKB}) holds as long as $\frac{S_p}{\hbar} \gg 1$.

The cyclic permutation operators can be expressed in terms of the spin
operators using the Dirac identity $P_{12}=\frac{1}{2}+2{\bf
S}_1\cdot{\bf S}_2$ and the spin Hamiltonian is
\begin{equation}
H=J_{nn}\sum_{nn}{\bf S}_i\cdot{\bf S}_j+J_{nnn}\sum_{nnn}{\bf
S}_i\cdot{\bf S}_j+\cdots \label{spinH}
\end{equation}
The first term in Eq.~(\ref{spinH}) is sum over distinct nearest
neighbors, the second is over distinct next nearest neighbors, and so
on. Here, $J_{nn}=4J_2+5J_4-4J_3+\cdots $ and $J_{nnn}=J_4+\cdots$. In
general, this is a highly competing magnetic Hamiltonian. On a regular
triangular lattice a model containing exchanges upto $J_5$ has been
studied by various approximate analytical and numerical finite size
(maximum of 36 sites) diagonalization methods.\cite{Misguich,Misguich1} The
picture that has emerged is rather complex containing a number of
broken symmetry states: a ferromagnetic, a three sublattice N{\'e}el,
a four sublattice N{\'e}el, and a long wavelength spiral states. In
addition, on the basis of numerical work, it has been argued that a
sizeable region of the phase diagram consists of a spin liquid state,
with short ranged correlations, spin gap, and no broken translational
and spin rotational symmetries.

\subsection{Magnetism: disordered system}
\label{sec: magdis}

In the presence of disorder, the picture should change substantially.
The system is no longer described by a regular triangular lattice and
will instead distort into a random lattice, with the sites dominantly
determined by the pinning defects. Those properties of the pure system
that are specific to a triangular lattice will no longer hold. For
example, none of the antiferromagnetic states, which depend delicately
on the regular lattice structure can be the true ground states.  More
fundamentally, there is no longer an argument that 3-particle exchange
is larger than the 2-particle exchange, rather the opposite could
hold, as we shall see.  To explore the effect of disorder, we
calculate the multiparticle exchange processes in a disordered system
whose low energy magnetic Hamiltonian can be formulated as in the pure
system but with a random distribution of exchange constants.

Leaving aside the gauge symmetry, the symmetries that are allowed to
be broken in a disordered system are the spin rotational
invariance ($\cal S$) and the time reversal invariance ($\cal T$). The
phases that are potentially important are a $\cal T$-broken metal, a
$\cal T$-broken insulator, a ${\cal S}$ and $\cal T$ broken
spin-glass, a disordered ferromagnet, and a disordered
antiferromagnet. Since a disordered system does not respect
translational invariance, no further subclassification according to
broken translational symmetry is possible. It is clear, however, that
the regime close to the crystalline phase of the pure system will be
marked by strong short-ranged crystalline order. Generically, disorder
necessarily renders all quantum phase transitions between these states
continuous, and thus the phase diagram is rife with quantum critical
points and lines.

\subsection{The Model}

\newcommand{\nimp}{N_{imp}} \newcommand{\gradient}{{\vec{\nabla }}}
\newcommand{\rvec}{{\mathbf{r}}} \newcommand{\pvec}{\rvec^{imp}}
\newcommand{\Svec}{{\mathbf{S}}} \newcommand{\momentum}{{\mathbf{p}}}

\begin{figure}
  \centerline{\epsfxsize=3.4in \epsffile{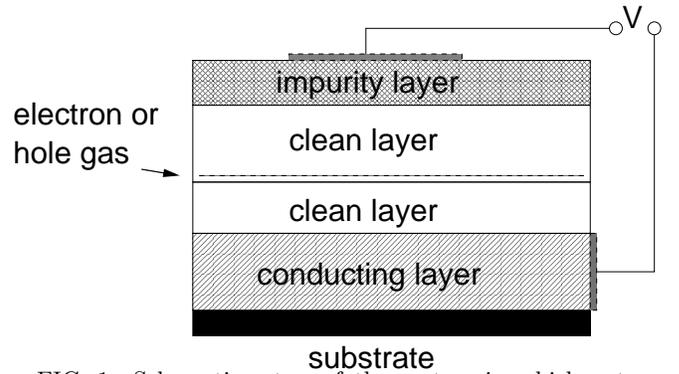}}
  \caption{Schematic setup of the system in which a two- dimensional
  electron or hole gas is generated.}  \label{fig: Setup}
\end{figure}

The systems of experimental interest differ considerably, but they can
be schematized as shown in Fig.~\ref{fig: Setup}.
The carriers themselves are confined to an inversion layer or a
quantum well with a width of the order of \( \sim 100 \)\AA. A buffer
of several hundred \AA\ separates the carrier plane from the doping
layer, which contains impurities in the form of oppositely charged
ions that provide the carriers.  We will use the language appropriate
to the electron-doped case for the sake of clarity; the hole-doped
case can be treated identically.

Let us denote the coordinates of the \( N \) electrons by \( \rvec_{i}
\), and those of the \( \nimp \) positively charged impurities by \(
\pvec_{j} \).  We will treat the carriers as being exactly confined to
the xy plane, so that \( \rvec_{i}=(x_{i},y_{i},0) \), which means
that we neglect the finite spread of the wavefunction in the direction
perpendicular to the plane. This spread leads to a softening of the
Coulomb potential at distances comparable with or smaller than the
effective Bohr radius \( a_{B} \). In the dilute limit consider here,
$r_s\gg 1$, the
many-particle wavefunction will be negligibly small in those regions
of coordinate space where two or more electrons come close enough to
each other to ``feel'' this softer potential.

We assume the only source of disorder is provided by the impurity ions
in the doping layer, which is separated by a distance \( d \) from the
carriers. We will consider the following model for the impurity
distribution: the thickness of the impurity layer is taken to be zero,
so that the impurities are exactly confined to the plane \( z=d \).
We have also considered a second model in which we assumed the
impurity layer to have a finite thickness, taken to be equal to the
separation \( d \) from the electron gas.  Since the results are very
similar, we shall not report them here.

Within the doping layer the impurities are randomly distributed. The
Hamiltonian is
\begin{equation}
H=\sum _{i=1}^{N}\left\{ \frac{\momentum_i^{2}}{2\meff } + \sum
  _{j=1}^{i-1}v\left( \rvec _{i}-\rvec _{j}\right) -\sum _{j=1}^{\nimp
    }v\left( \rvec _{i}-\pvec _{j}\right) \right\} ,
\label{eq:ham}
\end{equation}
where
\begin{equation}
v\left( \rvec \right) =\frac{e^{2}}{\epsilon }\frac{1}{|\rvec |}
\end{equation}
is the effective Coulomb potential, \( \epsilon \) being the dielectic
constant of the environment and \( \meff \) the effective mass of the
carriers.

\section{The Multiparticle Exchange Picture}


\newcommand{\Rvec}{{\mathbf{R}}}
\newcommand{\svec}{\underline{\sigma}}
\newcommand{\ket}[1]{\left| #1 \right\rangle }
\newcommand{\brakket}[3]{\left\langle #1 \right| #2 \left| #3
  \right\rangle } \newcommand{\bracket}[2]{\left\langle #1 |#2
  \right\rangle } \newcommand{\dr}{{\mathbf{u}}} \newcommand{\dt}{T}
\newcommand{\kroneckersigma}{\delta_{\svec_1\svec_2}}

It is useful to define the collective spatial and spin coordinates
\begin{equation}
\Rvec =\left( \rvec _{1},\rvec _{2},\cdots ,\rvec _{N}\right) ,\qquad
\svec =\left( \sigma _{1},\sigma _{2},\cdots ,\sigma _{N}\right) .\end{equation}
Formally we can view \( \Rvec \) as the coordinate of a single
particle moving in a \( 2N \)-dimensional space in the potential
$V(\Rvec )$; see Eq.~\ref{eq:ham}.

For Fermions, the
partition function of the system is then
\begin{equation}
\label{eqn_PartitionFunction}
  Z=\sum _{P\in S_{N}}(-1)^{P}\sum _{\svec }\int d\Rvec \,
    G(\Rvec ,\svec ;P\Rvec ,P\svec ;\beta )
\end{equation}
where the first sum is over all \( N! \) permutations \( P\Rvec \) of
the electron coordinates, and \( (-1)^{P} \) is the sign of the
permutation. The propagator is defined as
\begin{equation}
\label{eqn_Propagator}
  G(\Rvec _{1},\svec _{1};\Rvec _{2},\svec _{2};\tau )
  = \kroneckersigma
  \brakket{\Rvec _{1},\svec _{1}}{e^{-\tau H}}{\Rvec _{2},\svec _{2}}.
\end{equation}
Here \( \kroneckersigma \) is a product of \( N \) Kronecker delta
symbols. Note that this definition of the propagator treats the
electrons as distinguishable Boltzmann particles. Fermi statistics
have been taken into account in the sum over boundary conditions in
the partition function (\ref{eqn_PartitionFunction}).

\subsection{The Semiclassical Approximation}

\newcommand{\Pref}[1]{F\left[ #1 \right] }
\newcommand{\uvec}{{\mathbf{u}}}

The instanton method that we shall follow has been elegantly discussed by
Coleman.\cite{Coleman} The imaginary time path integral for the
propagator is ($T_{\tau}$ here denotes imaginary time)
\begin{equation}
\label{eqn_PIPropagator}
  G(\Rvec _{1},\svec _{1};\Rvec _{2},\svec _{2};T_{\tau} ) =\kroneckersigma
  \int _{\Rvec (0)=\Rvec _{1}}^{\Rvec (T_{\tau}) =\Rvec _{2}}D\Rvec
  \, e^{-\frac{1}{\hbar }S[\Rvec ]},
\end{equation}
where the Euclidean action is
\begin{equation}
\label{eqn_Action}
  S[\Rvec ]=\int _{0}^{T_{\tau} }d\tau \left\{ \frac{\meff }{2} \left(
  \frac{d\Rvec }{d\tau }\right) ^{2}+V(\Rvec )-V_{0}\right\} .
\end{equation}
The equilibrium potential energy \( V_{0}=\min _{\Rvec }V(\Rvec ) \)
has been subtracted out for later convenience. The stationary path
satisfies
\begin{equation}
\label{eqn_of_motion}
  \frac{\delta S[\Rvec _{c}]}{\delta \Rvec (\tau )} =-\meff
  \frac{d^{2}\Rvec _{c}}{d\tau ^{2}}+\gradient V(\Rvec _{c})=0,
\end{equation}
and the action for this path is
\begin{equation}
\label{eqn_ActionV}
  S[\Rvec _{c}]=\int _{\Rvec _{i}}^{\Rvec _{f}}dR\, \sqrt{2\meff
  \left( V(\Rvec )-V_{0}\right) }.
\end{equation}
The Planck constant enters the action in this form only through the
parameter \( r_{s}\sim 1/\hbar ^{2} \). Therefore, the semiclassical
caluclations described here are accurate in the low density limit,
$r_s\to \infty$.

The Gaussian quantum fluctuations around the stationary path are taken
into account by defining the fluctuation coordinates
\(
\dr (\tau )\equiv \Rvec (\tau )-\Rvec _{c}(\tau ) \), in terms of which we
expand the action to second order:
\begin{equation}
\int D\Rvec \, e^{-\frac{1}{\hbar }S[\Rvec ]} = F\left[ \Rvec
  _{c} \right] e^{-\frac{1}{\hbar }S\left[ \Rvec _{c} \right] },
\end{equation}
where
\begin{eqnarray}
\label{eqn_defPrefactor}
  F\left[ \Rvec _{c} \right] &=& \int D\dr \exp \left\{
  -\frac{1}{2\hbar } \int_{0}^{T_{\tau}}d\tau \, \dr (\tau )\cdot
  A(\tau )\dr (\tau )\right\} \nn &=& \left[ \det
  A\right]^{-1/2},
\end{eqnarray}
and we have assumed that the stationary path is unique. In cases where
more than one stationary path exists, their contributions have to be summed.
The differential operator \( A \) is given
for each path by
\begin{equation}
\label{eqn_DiffOperator}
  A_{\mu \nu }(\tau )=\left. -\delta _{\mu \nu }\meff
  \frac{d^{2}}{d\tau ^{2}}+\frac{\partial ^{2}V(\Rvec )}{\partial
  R_{\mu } \partial R_{\nu }}\right| _{\Rvec =\Rvec _{c}(\tau )}.
\end{equation}
The determinant is defined in terms of the eigenvalues \( \lambda
_{\nu } \) of \( A \), subject to the boundary conditions \(
\uvec (0)=\uvec (\dt )=0 \), as \( \det A = \prod _{\nu }\lambda
_{\nu } \).

\subsection{Exchange Processes and the Instanton Approximation}
\label{sec: Exchange}

\newcommand{\Rmin}{\overline{\Rvec }}
\newcommand{\Rinst}{\Rvec_{\mathrm{inst}}}
\newcommand{\Sinst}{S_{\mathrm{inst}}}
\newcommand{\attemptfreq}{\omega_{a}} \newcommand{\Om}{\hat{\Omega }}
\newcommand{\mand}{\quad \mbox {and}\quad } \newcommand{\Pone}{1}

We will assume that there exists a definite configuration \( \Rmin \)
of the \( N \) electrons that minimizes the electrostatic potential \(
V(\Rvec ) \).  It is clear that this classical minimum is \(
N! \)-fold degenerate, since the potential energy is invariant under
any permutation of the electron coordinates. In the semiclassical
limit, configurations where \( \Rvec \) is in the vicinity of one of
these minima will contribute dominantly to the partition function. We
will therefore construct stationary paths that begin and end at one of
those minima. In particular, we define the instanton path \( \Rinst
(\tau ) \) between the two minima at \(
\Rmin _{1} \) and \( \Rmin _{2} \) such that
\begin{equation}
\label{eqn_BoundaryConditions}
  \Rinst (-\infty )=\Rmin _{1}\mand \Rinst (+\infty )=\Rmin _{2},
\end{equation}
and the equation of motion (\ref{eqn_of_motion}) is satisfied. In the
simplest case, that of \( \Rmin _{1}=\Rmin _{2} \), the instanton path
is given by \( \Rinst (\tau )\equiv \Rmin _{1} \). In general \( \Rmin
_{1} \) and \( \Rmin _{2} \) differ by a permutation of the electron
coordinates \( \rvec _{i} \), so that the instanton path describes a
multiparticle exchange process.

In the vicinity of a minimum,  \( \Rmin \), 
\begin{equation}
\Rvec (\tau )-\Rmin \sim e^{\pm \tau \Om (\Rmin )}\uvec _{0},
\end{equation} 
where \( \uvec _{0} \) is some constant vector, and \( \Om \) is
defined as the square root of the Hessian matrix, evaluated at \(
\Rmin \):
\begin{equation}
\label{eqn_Hessian}
  \Omega ^{2}_{\mu \nu }(\Rmin )=\frac{1}{\meff }\left. \frac{\partial
  ^{2}V(\Rvec )}{\partial R_{\mu }\partial R_{\nu }} \right| _{\Rvec
  =\Rmin }.
\end{equation}
Hence any deviations from the classical equilibrium configuration are
localized on the imaginary time axis on a scale \( \delta \tau
\sim \frac{1}{\attemptfreq } \), where \( \attemptfreq ^{2} \) is some
eigenvalue (not necessarily the smallest) of \( \Om (\Rmin ) \). In this sense, 
the instanton path will be localized around the location of the instanton,
\( \tau_{\mathrm{inst}} \), in imaginary time. On a coarse-grained
time scale the exchange processes will therefore appear as
instantaneous, independend events.

The instanton path will be unique in most cases. An exception is the
two-particle exchange, where the electrons can take two equivalent
paths corresponding to clockwise and counterclockwise exchange. This
merely results in a factor of two for the exchange frequency.

The instanton formalism rests on the assumption, occasionally referred
to as the dilute gas approximation, that the average distance \(
\Delta \tau \) on the imaginary time axis between exchange processes
within the same region of space exceeds the instanton duration \(
\delta \tau \) by several orders of magnitude. If we consider the
propagator on a time scale \( T_{\tau} \) that satisfies
\begin{equation}
\label{eqn_tauLimits}
  \delta \tau \ll T_{\tau} \ll \Delta \tau
\end{equation}
we can make the following two crucial assumptions:

\begin{enumerate}
\item Each time slice contains at most one instanton event. Processes
  with two or more instanton events in a single time slice are of
  second order in \( T_{\tau} /\Delta \tau \) and therefore negligible.
\item Instantons do not occur within a few instanton lengths of a time
  slice boundary.  Again, processes that violate this assumption are
  of order \( (T_{\tau} /\Delta \tau )(\delta \tau /T_{\tau} )=\delta \tau
  /\Delta \tau \) and therefore negligible.
\end{enumerate}
Let us now evaluate the propagator within these approximations. Since
the Hamiltonian is independent of spin we will suppress the spin
indices in our notation for the moment. We also define a fluctuation
coordinate \( \dr =\Rvec -\Rmin \), where \( \Rmin \) is by definition
the particular minimum of \( V(\Rvec ) \) that is closest to \( \Rvec
\). Thus we want to evaluate
\begin{equation}
\label{eqn_ThePropagator}
  G(\Rmin _{1}+\dr _{1};\Rmin _{2}+\dr _{2};T_{\tau} )=
    \int _{\Rvec (0)=\Rmin _{1}+\dr _{1}}^{\Rvec (\tau _{1})=
    \Rmin _{2}+\dr _{2}}D\Rvec \, e^{-\frac{1}{\hbar }S[\Rvec ]},
\end{equation}
where the deviations \( \dr _{1} \) and \( \dr _{2} \) are by
assumption 2 in the quadratic regime, so that we can expand the
equation of motion to linear order in \( \uvec \). In other words, we
are allowed to approximate
\begin{equation}
\label{eqn_ApproximateOmega}
  \Om (\Rmin +\uvec )\simeq \Om (\Rmin ).
\end{equation}

An approximate solution of the equation of motion
(\ref{eqn_of_motion}) that satisfies the boundary conditions
\begin{equation}
\Rvec _{c}(0)=\Rmin _{1}+\uvec _{1} \mand \Rvec _{c}(T_{\tau} )=\Rmin
_{2}+\uvec _{2}\end{equation} is then

\begin{equation}
\label{eqn_approximatePath}
  \Rvec _{c}(\tau )=e^{-\tau \Omega _{1}^{1/2}}\dr _{1} +e^{(\tau -T_{\tau}
  )\Omega _{2}^{1/2}}\dr _{2}+\Rinst (\tau -\tau _{0}),
\end{equation}
where \( \tau _{0} \) is an arbitrary reference point between \( 0 \)
and \( \tau _{1} \). The time derivative of each term is localized on
a time scale \( \delta \tau \ll T_{\tau} \), and by assumption 2 above the
overlap between the three terms is exponentially small. Hence the
corrections arising from the nonlinearity of the equation of motion
are negligible. For the same reason the action associated with this
path splits into three parts, which we write in an obvious notation as
\begin{equation}
S[\Rvec _{c}]=S[\uvec _{1}]+S[\uvec _{2}]+\Sinst .\end{equation}

With the results of the previous section the propagator is then
\begin{eqnarray}
& & G(\Rmin _{1}+\dr _{1};\Rmin _{2}+\dr _{2};T_{\tau} ) \nn
& & =\Pref{\Rinst }\exp \left\{
  -\frac{1}{\hbar }\left( S[\dr _{1}]+S[\dr _{2}]+\Sinst \right) \right\} ,
\end{eqnarray}
where we already incorporated the fact that with the approximation
(\ref{eqn_ApproximateOmega}) the prefactor (\ref{eqn_defPrefactor}) is
independent of the \( \uvec _{i} \).  Let us from now on write \(
\Rinst =\Rvec _{P} \) and \( \Sinst =S_{P} \), where \( P\in S_{N} \)
labels the particular permutation that takes \( \Rmin _{2} \) into \(
\Rmin _{1} \), i.e. \( \Rmin _{1}=P\Rmin _{2} \). For later
use we define the quantity \( G_{P} \) as the ratio between the propagator for a
given instanton path and the corresponding propagator for the trivial
path \( \Rinst (\tau )\equiv \Rmin \), which has \( \Sinst =0 \). That
is,
\begin{equation}
\label{eqn_GP}
  G_{P}:=\frac{G(\Rmin +\dr _{1};P\Rmin +P\dr _{2};T_{\tau} )}
              {G(\Rmin +\dr _{1};\Rmin +\dr _{2};T_{\tau} )}
        =\frac{\Pref{\Rvec _{P}}}{\Pref{\Rmin }}e^{-\frac{1}{\hbar }S_{P}}.
\end{equation}

Within the instanton approximation \( G_{P} \) is independent of the
fluctuation coordinates. Due to permutation symmetry all terms in \(
G_{P} \) are also independent of the particular choice of the minimum
\( \Rmin \).

\subsection{The Prefactor}
\label{sec: Prefactor}

\newcommand{\detprime}{\mathrm{det}'}

To evaluate the prefactor (\ref{eqn_defPrefactor}) we would have to
find a complete set of eigenfunctions \( \uvec _{n}(\tau ) \) that
satisfy
\begin{equation}
\label{eqn_Eigenvalues}
  \left[ -\meff \delta _{\mu \nu }\partial _{\tau }^{2}
    +V_{\mu \nu }(\tau )\right] u_{n\nu }(\tau )=\lambda _{n}u_{n\mu }(\tau )
\end{equation}
with the boundary conditions \( \dr _{n}(0)=\dr _{n}(T_{\tau})=0 \).
Here we used the shorthand notations \( \partial _{\tau
  }^{2}=d^{2}/d\tau ^{2} \) and \( V_{\mu \nu }(\tau )=\partial
^{2}V(\Rvec _{P}(\tau ))/\partial R_{\mu }\partial R_{\nu } \).  If we
expand
\begin{equation}
\label{eqn_EigenfuncExpansion}
  \dr (\tau )=\sum _{n=0}^{\infty }c_{n}\dr _{n}(\tau )
\end{equation}
the path integral over \( \uvec \) is transformed into
\begin{equation}
\int D\uvec \to \prod _{n=0}^{\infty } \int \frac{dc_{n}}{\sqrt{2\pi
      \hbar }} .\end{equation}

We still have to account for the possibility that one of the
eigenvalues of (\ref{eqn_DiffOperator}) is less than or equal to zero.
While it is easy to show by direct calculation that this is not the
case for \( \Pref{\Rmin } \), a zero eigenvalue indeed exists for \(
\Pref{\Rvec _{P}} \). As an eigenfunction we consider the time
derivative of the instanton path itself:
\begin{equation}
\dr _{0}(\tau ):=a_{0}^{-1}\frac{d}{d\tau }\Rvec _{P}(\tau -\tau
_{0}),\end{equation} where the normalization constant is given by
\begin{equation}
a_{0}^{2}=\int _{0}^{T_{\tau} }d\tau \left[ \frac{d}{d\tau }\Rvec _{P}(\tau
  -\tau _{0})\right] ^{2}=\frac{S_{P}}{\meff }.\end{equation}

The last identity follows from the equation of motion
(\ref{eqn_of_motion}), which can be integrated to give
\begin{equation}
\label{eqn_EnergyConservation}
  \frac{\meff }{2}\left( \frac{d\Rvec _{c}}{d\tau }\right) ^{2}
   = V(\Rvec _{c})-V_{0},
\end{equation}
which is just the Euclidean version of energy conservation. It is
straightforward to verify that \( \dr _{0}(\tau ) \) satisfies the
eigenvalue equation (\ref{eqn_Eigenvalues}) with eigenvalue \( \lambda
_{0}=0 \). The boundary conditions \( \uvec (0)=\uvec (T_{\tau} )=0 \) are
satisfied within our approximations since \( \uvec _{0} \) is
exponentially localized. This function just describes the change in \(
\Rvec _{c} \) due to a change in the instanton position \( \tau _{0}
\), which is arbitrary within the limits \( 0<\tau _{0}<T_{\tau} \). Hence
a shift in the instanton position corresponds to a zero mode. This is
the Goldstone mode associated with broken time translation
symmetry in the presence of an instanton. The change in the path \(
\Rvec (\tau )=\Rvec _{P}(\tau )+\dr (\tau ) \) due to a change in the
expansion coefficient \( c_{0} \) can be related to a change in \(
\tau _{0} \) as follows:
\begin{equation}
\frac{d\Rvec (\tau )}{dc_{0}}=\uvec _{0}(\tau ),\end{equation} by
Eq.~(\ref{eqn_EigenfuncExpansion}), while
\begin{equation}
\frac{d\Rvec (\tau )}{d\tau _{0}}=-\frac{d\Rvec _{P}(\tau -\tau
  _{0})}{d\tau }=-a_{0}\uvec _{0}(\tau )\end{equation} by the definition of \(
\uvec _{0} \), so that we have to replace the integration over \(
c_{0} \) by
\begin{equation}
\int \frac{dc_{0}}{\sqrt{2\pi \hbar }}\to a_{0}\int _{0}^{T_{\tau}
  }\frac{d\tau _{0}}{\sqrt{2\pi \hbar }}=T_{\tau} \sqrt{\frac{S_{P}}{2\pi
    \hbar \meff }}.\end{equation}

\( \lambda _{0} \) is the lowest eigenvalue, since the corresponding
eigenfunction is free of nodes. Hence all other eigenvalues must be
positive.  We now have
\begin{equation}
F[\Rvec _{P}]=T_{\tau} \sqrt{\frac{S_{P}}{2\pi \hbar \meff }}\left(
  \detprime \left[ -\meff \partial _{\tau }^{2}+V_{\mu \nu }(\tau
    )\right] \right) ^{-1/2},\end{equation} where the prime indicates that the
zero eigenvalue has to be omitted in the determinant. To summarize,
the prefactor is given by
\begin{equation}
\frac{\Pref{\Rvec _{P}}}{\Pref{\Rmin }}=T_{\tau} \sqrt{\frac{S_{P}}{2\pi
    \hbar \meff }}\left( \frac{\det \left[ -\meff \partial _{\tau
        }^{2}+V_{\mu \nu }(0)\right] }{\detprime \left[ -\meff
      \partial _{\tau }^{2}+V_{\mu \nu }(\tau )\right] }\right)
^{1/2}.\end{equation}

Let us assume that we scale all eigenmodes of the potential by the
same factor \( g \) and simultaneously rescale the imaginary time
variable by a factor \( g^{-1} \).
The factors of \( g \) in the determinants cancel in the numerator and
in the denominator for each eigenvalue separately, and we know that
the ratio of determinants cannot depend on the length \( T_{\tau} \) of the
time interval, except for exponentially small corrections. Hence the
prefactor depends linearly on a characteristic frequency scale of \(
V(\Rvec ) \), and the ratio of determinants depends only on the
relative values of the eigenfrequencies.  We summarize these findings
by writing \( G_{P} \), defined in Eq.~(\ref{eqn_GP}), as
\begin{equation}
\label{eqn_GP_explicit}
  G_{P}=T_{\tau} A_{P} \, \omega_0 \, \sqrt{\frac{S_{P}}{2\pi \hbar }}
        \; e^{-\frac{1}{\hbar }S_{P}},
\end{equation}
where, as stated above, \( \omega_0 \) is a characteristic
frequency, and the dimensionless factor \( A_{P} \) depends on the
relative values of the eigenfrequencies during the exchange process.
It seems reasonable to assume that \( A_{P} \) is roughly of order
one. Although the prefactor is not expected to cause any drastic
changes in our results, it is still interesting to determine the
change in characteristic frequency with disorder. It is conceivable,
for example, that disorder would bring about a reduction in the phonon
spectrum, and this mechanism could lead to a suppression of exchange
processes.

\subsection{The Exchange Hamiltonian}

\newcommand{\TM}{\hat{T}} \newcommand{\Ps}{\hat{P}^{\sigma }}

Our goal in this section is a Hamiltonian description of the system in
terms of multiparticle exchange operators.  In the previous
subsections we calculated the imaginary-time propagator on an
intermediate time scale $T_{\tau}$ defined by the relation
(\ref{eqn_tauLimits}). To apply this result, we split the partition
function (\ref{eqn_PartitionFunction}) into \( M \) imaginary time
slices, where $M$ satisfies \( \beta = M T_{\tau} \). This requires us
to sum over $M-1$ intermediate configurations, so that the partition
function reads
\begin{eqnarray}
  Z & = & \sum_{P}(-1)^{P}\sum _{\svec_{1}} \cdots \sum_{\svec _{M}}
          \int d\Rvec_{1}\cdots \int d\Rvec_{M} \nn
 &\times& G(\Rvec _{1},\svec _{1};\Rvec _{2},\svec _{2};T_{\tau} ) \nn
      & & \cdots G(\Rvec _{M},\svec _{M};P\Rvec _{1},P\svec_{1};T_{\tau} ).
\label{eqn_PartitionSlices}
\end{eqnarray}

We want to make use of the quantities
\begin{equation}
\label{eq_GP_again}
  G_{P} =\frac{G(\Rmin +\dr _{1};P\Rmin +P\dr _{2};T_{\tau} )}
              {G(\Rmin +\dr _{1};\Rmin +\dr _{2};T_{\tau} )} ,
\end{equation}
defined in Sec.~\ref{sec: Exchange}, which only depend on the
permutation $P$, and are independent of the fluctuation coordinates
$\dr_i$ and the particular choice of the minimum $\Rmin$.  To this
end, we write the integration variables $\Rvec_i$ in the form
\begin{equation}
  \Rvec_i = P_i(\Rmin+\dr_i),
\end{equation}
where $\Rmin$ is some minimum of $V(\Rvec)$, and the permutation $P_i$
is chosen such as to minimize the distance $|\Rvec_i - P_i\Rmin|$.
The integrals then
have to be replaced with \begin{equation}   \int d\Rvec_i \to \sum
_{P_i} \int d\dr_i, \end{equation}
where the sum is over all permutations, so that $P_i \Rmin$ covers all
minima of $V(\Rvec)$, and the integration over $\dr_i$ is by
construction restricted to the vicinity of $\dr_i=0$. The partition
function then reads (dropping spin in the notation for now)
\begin{eqnarray}
\label{eq_partition_new}
  Z & = & \sum_{P}(-1)^{P} \sum_{P_1} \cdots \sum_{P_M}
          \int d\dr_1 \cdots \int d\dr_M \\
 &\times& G\Bigl( P_1(\Rmin+\dr_1); P_2(\Rmin+\dr_2); T_{\tau} \Bigr) \nn
 & &       \cdots G\Bigl( P_M(\Rmin+\dr_M); P P_1(\Rmin+\dr_1); T_{\tau}
\Bigr). \nonumber
\end{eqnarray}

Let us now introduce the transfer matrix $\TM$, defined by the
relation
\begin{eqnarray}
\label{eq_transfermatrix}
  && G\Bigl( P_i(\Rmin+\dr_i), \svec_i; P_j(\Rmin+\dr_j), \svec_j; T_{\tau}
\Bigr) \nn
  && = \brakket{i,\svec_i}{\TM }{j,\svec_j}\,
G(\Rmin+\dr_i;\Rmin+\dr_j;T_{\tau} ).
\end{eqnarray}
Comparing this definition to Eq.~(\ref{eq_GP_again}) we can easily
deduce
\begin{equation}
  \brakket{i,\svec_i}{\TM }{j,\svec_j} = \kroneckersigma G_{P_{ij}} =
\sum_P G_P \brakket{i,\svec_i}{ \hat P' }{j,\svec_j},
\end{equation}
where $P_{ij} = P_i^{-1}P_j$,  the permutation operators \( \hat{P}'
\) are defined to act only on the index \( i \) as \(
\hat{P}'\ket{i,\svec }=\ket{p_{i},\svec } \), and we made use of the
orthogonality relation $\bracket{i,\svec_i}{j,\svec_j} =
\delta_{ij}\kroneckersigma$. Thus the transfer matrix is
\begin{equation}   \TM = \sum_P G_P \hat P'. \end{equation}

Inserting Eq.~(\ref{eq_transfermatrix}) into the partition function
(\ref{eq_partition_new}), the latter will factorize into a fluctuation
part and a tunneling part:
\begin{eqnarray}
  Z & = & Z_0 \sum_{P}(-1)^{P} \sum_{i_1} \cdots \sum_{i_M}
\sum_{\svec_1} \cdots \sum_{\svec_M} \nn
&& \times \brakket{i_1,\svec_1}{\TM}{i_2,\svec_2} \cdots
\brakket{i_M,\svec_M}{\TM \hat P}{i_1,\svec_1} \nn
     & = & Z_0 \sum_P(-1)^P \sum_{i\svec}
\brakket{i,\svec}{\TM^M \hat P }{i,\svec} \nonumber,
\end{eqnarray}
where the permutation operator \( \hat{P} \) acts on both
$i$ and $\svec_i$ as \( \hat{P}\ket{i,\svec }=\ket{p_{i},P\svec } \),
and
\begin{eqnarray}
Z_{0} & = & \int d\dr _{1}\cdots \int d\dr _{M}\,
            G(\Rmin+\uvec_1;\Rmin+\uvec_2;T_{\tau} ) \nn
      &   & \cdots G(\Rmin+\uvec_M;\Rmin+\uvec_1;T_{\tau} )
\end{eqnarray}
is the partition function for a \( 2N \)-dimensional harmonic
oscillator.

The \( G_{P} \) are proportional to the lenght of a time slice \( T_{\tau}
= \beta/M \), see Eq.~(\ref{eqn_GP_explicit}), with the
exception of the identical permutation \( P=1 \), for which \( G_{1}=1
\). We therefore define the exchange energies \( J_{P}=\frac{M}{\beta
  }G_{P} \), which allows us to write
\begin{equation}
\hat{T}^{M}=\left[ 1+\frac{\beta }{M}\sum _{P\ne
    1}J_{P}\hat{P}'\right] ^{M}=\exp \left\{ \beta \sum _{P\ne
    1}J_{P}\hat{P}'\right\} \end{equation}
in the zero-temperature limit, in which
\( M=\beta/T_{\tau} \to \infty \). The partition function
now reads

\begin{equation}
Z=Z_{0}\sum _{P}(-1)^{P}\sum _{i\svec }\brakket{i,\svec }{\exp \left\{
    \beta \sum _{P'\ne 1}J_{P'}\hat{P}'\right\} \, \hat{P}}{i,\svec
  }.\end{equation}

This is the desired representation in terms of permutation operators.
The exchange energies are given by Eq.~(\ref{eqn_GP_explicit}) as
\begin{equation}
\label{eqn_ExchangeFrequencies}
  J_{P}=A_{P}\, \hbar \omega _{a}\sqrt{\frac{S_{P}}{2\pi \hbar }}
       \: e^{-\frac{1}{\hbar }S_{P}}.
\end{equation}

If we expand the exponential in a power series,
the orthogonality condition \( \bracket{i,\svec }{j,\svec }=\delta
_{ij} \) implies that all permutation operators \( \hat{P}' \) in this
expansion have to combine with \( P \) to the
identical permutation: \( \hat{P}_{1}'\hat{P}_{2}'\cdots
\hat{P}_{n}'\hat{P}\ket{i,\svec }=\ket{i,P\svec } \), or \(
P=(P_{n}')^{-1}\cdots (P_{1}')^{-1} \) as far as their action on \( i
\) is concerned. We can thus eliminate the sum over \( P \) and absorb
the spin permutations and the sign factor into the exponential. Then
the sum over \( i \) is redundant due to permutation symmetry and the
partition function becomes

\begin{equation}
Z=N!Z_{0}\sum _{\svec }\brakket{\svec }{\exp \left\{ -\beta \sum
    _{P\ne 1}(-1)^{P+1}J_{P}\Ps \right\} }{\svec },\end{equation} where \(
\hat{P}^{\sigma } \) acts on the spin variables as \( \hat{P}^{\sigma
  }\ket{\svec }=\ket{P\svec } \).  This is the partition function for
a pure spin Hamiltonian
\begin{equation}
\label{eqn_SpinHamiltonian}
H_{\sigma }=\sum _{P}(-1)^{P+1}J_{P}\Ps.
\end{equation}

\subsection{Generalized Heisenberg Model}

\newcommand{\Spin}{{\mathbf{S}}} \newcommand{\PP}{\hat{P}}

The spin permutation operators appearing in
Eq.~(\ref{eqn_SpinHamiltonian}) can be rewritten in terms of Pauli
spin operators. For example, if we denote by \( \Ps _{12} \) the
permutation operator that interchanges \( \sigma _{1} \) and \( \sigma
_{2} \),
\begin{eqnarray}
\Ps _{12} &=& \hat{\sigma }_{1}^{+}\hat{\sigma }_{2}^{-}+\hat{\sigma
 }_{1}^{-}\hat{\sigma }_{2}^{+}+\case{1}{2}\left( \hat{\sigma
   }_{1}^{z}\hat{\sigma }_{2}^{z}+1\right) \nn
   &=& 2\Spin _{1}\cdot \Spin_{2}+\case{1}{2},
\end{eqnarray}
leading to a Heisenberg term, as one can easily
check by direct calculation of the matrix elements. Any permutation
can be written as a combination of elementary transpositions, and
hence as a product of spin operators. In general these products can be
reduced using operator identities such as \cite{Misguich}
\begin{eqnarray}
 \Ps _{123}+\Ps _{321} & = & \Ps _{12}+\Ps _{23}+\Ps _{31}-1,\nn
 \Ps _{1234}+\Ps _{4321} & = & \Ps _{12}\Ps _{34}+\Ps _{14}\Ps _{23}-
             \Ps _{13}\Ps _{24} \nn
                         &   & + \Ps _{13}+\Ps _{24}-1,
\end{eqnarray}
etc.
Keeping only the dominant two-, three-, and four-particle exchange
processes, the spin Hamiltonian becomes
\begin{eqnarray}
 H & = & (2J_{2}-4J_{3}+2J_{4})\sum _{<ij>}^{NN}\Spin _{i}\cdot
         \Spin _{j}+2J_{4}\sum _{<ij>}^{NNN}\Spin _{i}\cdot \Spin
_{j}\nn    &  & +4J_{4}\sum _{<ijkl>}^{\diamond }\left(
G_{ijkl}+G_{iljk}-G_{ikjl}         \right) ,
\end{eqnarray}
where \( NN \) indicates a sum over nearest-neighbor pairs, \( NNN \)
a sum over next-nearest neighbors, and \( \diamond \) is a sum over
all rhombi.  \( G_{ijkl}=(\Spin _{i}\cdot \Spin _{j})(\Spin _{k}\cdot
\Spin _{l}) \), where the vertices of the rhombus are labeled
clockwise by the four indices. This Hamiltonian has been discussed
in Secs.~\ref{sec: magpure} and \ref{sec: magdis}.

\section{Numerical Techniques}

\newcommand{\Xvec}{{\mathbf{X}}}
\newcommand{\Xmin}{{\overline{\Xvec}}}
\newcommand{\xvec}{{\mathbf{x}}} \newcommand{\yvec}{{\mathbf{y}}}

\newcommand{\nslices}{M}
\newcommand{\nmobile}{{N_{\text{mobile}}}}

\subsection{Calculation of the Action}
\label{sec: Numerics}

The action (\ref{eqn_ActionV})
\begin{equation}
S_{P}=\int _{\Rmin }^{P\Rmin }dR\sqrt{2\meff \left( V(\Rvec
    _{P})-V_{0}\right) }\end{equation} depends only on a single length scale,
which can be factored out. We define a dimensionless coordinate \(
\Xvec =\frac{1}{a}\Rvec \), where \( a \) is the lattice constant of
the ordered Wigner crystal. The unit cell of the triangular lattice is
of area \( A=\frac{\sqrt{3}}{2}a^{2} \), so that the density is
\begin{equation}
n_{s}=\frac{1}{A}=\frac{2}{\sqrt{3}}a^{-2},\end{equation} which we use as a
definition for \( a \) in the presence of disorder. The parameter \(
r_{s} \), defined in Sec.~\ref{sec: WignerCystal}, can be expressed in
terms of \( a \) as
\begin{equation}
r_{s}= \frac{3^{1/4}}{\sqrt{2\pi}} \frac{a}{a_{B}}\simeq
0.525\frac{a}{a_{B}}.
\end{equation}

We define the dimensionless action \( \tilde{S}_{P} \) by \(
\frac{1}{\hbar }S_{P}=r_{s}^{1/2}\tilde{S}_{P}. \) Then
\begin{equation}
\label{eqn_dimless_action}
  \tilde{S}_{P}=\eta \int _{\Xmin }^{P\Xmin }dX
                \sqrt{\tilde{V}(\Xvec _{P})-\tilde{V}_{0}},
\end{equation}
where
\begin{equation}
\label{eqn_DimlessPotential}
  \tilde{V}(\Xvec )=\sum _{i=1}^{N}\left\{ \sum _{j=1}^{i-1}
                    \frac{1}{\left| \xvec _{i}-\xvec _{j}\right| }
                   -\sum _{j=1}^{\nimp }\frac{1}{\left| \xvec _{i}
                   -\xvec ^{imp}_{j}\right| }\right\}
\end{equation}
is the dimensionless potential, \( \tilde{V}_{0} \) is its minimum
value, and \( \eta \) is a numerical factor:
\begin{equation}
\eta = \sqrt{2}\left( \frac{2\pi}{\sqrt{3}} \right)^{1/4}
\simeq 1.952.
\end{equation}

The classical path that minimizes the action has to be found
numerically.  Therefore we discretize the integral in
(\ref{eqn_dimless_action}) using the trapezoidal rule, which leads to
\begin{eqnarray}
\label{eqn_discreteaction}
\tilde{S} &\simeq& \frac{\eta }{2}\sum _{i=0}^{\nslices-1}\left| \Xvec _{i+1}-\Xvec
  _{i}\right| \nn
  & & \times \left\{ \sqrt{\tilde{V}(\Xvec
    _{i})-\tilde{V}_{0}}+\sqrt{\tilde{V}(\Xvec _{i+1})-\tilde{V}_{0}}
  \right\}.
\end{eqnarray}

The displacement of the participating electrons from their equilibrium
positions creates dipole perturbations, which are screened out after a
distance of a few lattice spacings, even in the absence of
conventional screening. We can therefore restrict the number of moving
particles to a relatively small value \( \nmobile \) and hold all other
particle coordinates fixed at their equilibrium values. Details on the
errors due to the finite values of \( \nslices \) and \( \nmobile \) 
can be found in Appendix~\ref{sec_accuracy}. Since these errors are of
opposite sign, we believe that the total error for the action is no
larger than \( 0.3\% \) in the clean system.  In order to keep the
distances \( \left| \Xvec   _{i+1}-\Xvec _{i}\right| \) approximately
constant during the minimization process, the allowed variations in \(
\Xvec _{i} \) are restricted to those satisfying \( \delta \Xvec
_{i}\cdot \left( P\Xmin -\Xmin \right) =0 \), thereby reducing the
number of independent variables per time slice by one. Since initial \(
(i=0) \) and final \( (i=\nslices) \) conditions are held fixed, the
action is a function of \( (2N-1)(M-1) \) independent variables in its
discretized form. For calculations on the clean system we took \(
\nslices=16 \) and \( \nmobile \simeq 80 \), depending on the particular
exchange under consideration. The minimization thus involves
around \( 2400 \) variables. We used a variable metric (quasi-Newton)
algorithm. \cite{NumericalRecipes} Due to the long range nature of the
Coulomb potential the sum in the expression (\ref{eqn_DimlessPotential})
for the potential energy converges very slowly, and is in fact only
conditionally convergent. We therefore use the Ewald summation
technique, in which the summation over the
long-range part of the Coulomb potential is carried out in Fourier
space. To improve the speed of the computation, we tabulated the Ewald
summation formulas on a \( 50\times 50 \) grid and calculated in-between
values using bicubic interpolation. We explicitly checked that the
interpolation procedure does not generate any errors comparable to the
stated accuracy of the results. In this way a single minimization could
be carried out in less than 10 minutes CPU time on a 400MHz Pentium II
processor.

\subsection{The Prefactor}
\label{sec_num_prefactor}

\newcommand{\evec}{{\mathbf \hat{e}}}
\newcommand{\hessian}{H}

We now turn to the numerical evaluation of the prefactor
(\ref{eqn_defPrefactor}), which we write in the form
\begin{eqnarray}
\label{eq_newPrefactor}
  \Pref{\Rvec_c} &=& \int D\uvec \; e^{-S[\uvec]}, \\
  S[\uvec] &=& \frac{\meff}{2\hbar} \int_0^{T_{\tau}} d\tau \: \left({\dot
\uvec(\tau)}^2 + \uvec(\tau) \cdot H(\tau) \uvec(\tau)\right), \nonumber
\end{eqnarray}
where
\begin{equation}
  H_{\mu\nu}(\tau) = \frac{1}{\meff} \left. \frac{\partial^2
    V(\Rvec)}{\partial R_\mu \partial R_\nu} \right|_{\Rvec =
  \Rvec_c(\tau)}
\end{equation}
is the Hessian matrix of the potential for the configuration at time
$\tau$. We split the imaginary
time axis into $M$ intervals, so that
\begin{equation}
  T_{\tau} = s_0 > s_1 > \cdots > s_{M-1} > s_M = 0,
\end{equation}
and approximate $H(\tau)$ by a constant matrix $H_i$ within a given time
interval \(  s_i > \tau > s_{i+1} \):
\begin{equation}
  H_{\mu\nu}(\tau) \simeq (H_i)_{\mu\nu} = \frac{1}{\meff} \left.
    \frac{\partial^2 V(\Rvec)}{\partial R_\mu \partial R_\nu}
    \right|_{\Rvec =   \Rvec_i},
\end{equation}
where $\Rvec_i = a \Xvec_i$ are the points of the discretized instanton
path determined in Sec.~\ref{sec: Numerics}. The corresponding
times $s_i$ can be calculated by inverting the equation of motion:

\begin{eqnarray}
    s_{i+1}-s_i &=& \int
        dR \left[ \frac{2}{\meff} (V(\Rvec)-V_0) \right]^{-1/2} \\
    &\simeq& \left(\frac{\meff}{8}\right)^{1/2}
        \frac{ |\Rvec_{i+1}-\Rvec_i| + |\Rvec_i-\Rvec_{i-1}| }
    {\sqrt{ V(\Rvec_i)-V_0 }}.  \nonumber
\end{eqnarray}

We can then write the prefactor in the form
\begin{eqnarray}
  \Pref{\Rvec_c} &\simeq& \int d\uvec _{1}G_{1}(0,\uvec _{1};T_{\tau}-s_{1})
    \nn &\times& \int d\uvec _{2}G_{2}(\uvec _{1},\uvec_{2};s_{1}-s_{2})
    \nn && \cdots G_{M}(\uvec _{M-1},0;s_{M-1}),
\end{eqnarray}
where
\begin{eqnarray}
G_{i}(\uvec_i, \uvec_{i+1}; s)
&=& \int D\uvec \, \exp \left\{
  -\frac{\meff}{2\hbar} \int_0^s d\tau \right. \nn
& & \left. \left( \frac{d\uvec}{d\tau}^2
    +\uvec(\tau) \cdot H_{i} \, \uvec(\tau ) \right) \right\}
\end{eqnarray}
is simply the propagator of a multidimensional harmonic oscillator and
can easily be calculated. We define orthonormal eigenvectors $\evec^i_\nu$ and eigenvalues $\omega_{i\nu}^2$ satisfying
\begin{equation}
  H_i \, \evec^{i}_{\nu } = \omega_{i\nu}^2 \, \evec ^{i}_{\nu }
\end{equation}
(note that $\omega_{i\nu}$ can be imaginary), in terms of which
\begin{eqnarray}
&&G_{i}(\uvec_1, \uvec_2; s) =
  \left( \prod_\nu B_{i\nu}(s) \right)^{1/2} \times \\
&&
  \exp \left\{ -\sum_\nu B_{i\nu}(s)
    \left[ \frac{1}{2} (u_{\nu 1}^2 + u_{\nu 2}^2)\cosh \omega_\nu s
    - u_{\nu 1} u_{\nu 2} \right]
    \right\}, \nonumber
\end{eqnarray}
where $u_{\nu 1,2} = \evec^i_\nu \cdot \uvec_{1,2}$
and
\begin{equation}
  B_{i\nu}(s) = \frac{\meff\omega_{i\nu}}{\hbar \sinh \omega_{i\nu} s}.
\end{equation}

The prefactor is then
\begin{equation}
\Pref{\Rvec_c}=\left( \prod _{i\nu } B_{i\nu}(s_i - s_{i-1})
      \right) ^{-1/2}\left( \det M\right) ^{-1/2},
\end{equation}
where the matrix \( M \) is defined as
\begin{eqnarray}
  M_{\mu \nu }^{ij} & = & \delta _{i,j}\left( A^{i}_{\mu \nu }
    +A^{i-1}_{\mu \nu }\right) - \left( \delta _{i,j+1}B^{i}_{\mu \nu }
    +\delta _{i,j-1}B^{j}_{\mu \nu }\right) ,\nn
  A^{i}_{\mu \nu } & = & \sum _{\lambda }e^i_{\lambda \mu }e^i_{\lambda \nu }
    \frac{\meff\omega_{i\lambda }}{\tanh \omega _{i\lambda
    }(s_{i}-s_{i-1})},\nn
  B^{i}_{\mu \nu } & = & \sum _{\lambda
    }e^i_{\lambda \mu }e^i_{\lambda \nu }     \frac{\meff\omega
    _{i\lambda }}{\sinh \omega _{i\lambda }(s_{i}-s_{i-1})}.
\end{eqnarray}

Numerical evaluation of the determinant is now straightforward. The
eigenvalue that corresponds to the zero mode of Sec.~\ref{sec:
Prefactor} has to be omitted from the result (this eigenvalue
will not be exactly zero here, due to the finite number of time slices).
To this task, we replace \( H(\tau) \) by \( H(\tau) -\lambda \) in
Eq.~(\ref{eq_newPrefactor}), and numerically search for the smallest
value of \( \lambda \) that satisfies \( 1/F(\lambda )=0 \). We then
divide the determinant by this value. The method outlined here has been
tested on the problem of tunneling in a quartic potential in one
dimension, which can be treated analytically; details can be found in
Appendix~\ref{sec_doublewell}.

\subsection{The Disordered System}

For the disordered system we have to sample over a large
number of impurity distributions. After placing the impurities onto
random locations in systems with \( 48 \) to \( 280 \) particles and
periodic boundary conditions, we first minimize the potential energy
of the classical electron configuration.  No tabulation of the Ewald
summation formulas was used in this minimization, since the classical
equilibrium configuration is very sensitive to numerical errors. We
cannot exclude the possibility that the minimization procedure gets
trapped in a metastable configuration in the presence of strong
disorder. On average, however, the properties of such a metastable
state should be sufficiently similar to those of the true equilibrium
state that our results will not be affected. For strong disorder, when
the triangular lattice structure is compromised even on short length
scales, we are also faced with the problem of identifying proper sets
of nearest neighbors to participate in the exchange. This task is
solved by a Delaunay triangulation of the electrons' equilibrium
coordinates. For the subsequent minimization of the
discretized action only the \( \nmobile=32 \cdots 34 \) particles
closest to those participating in the exchange were allowed to move,
with the remaining particles held fixed at their equilibrium positions.
The number of time slices was reduced to \( \nslices=8 \), so that we
have to minimize over approximately \( 500 \) independent variables.
The minimization converges significantly slower than in the clean
system, since the dependence of the action on the independent variables
is less smooth. In the presence of strong disorder each minimization
takes several minutes to carry out. Typically we generated around \( 250
\) impurity configurations, for each of which \( 8 \) exchange processes
were chosen at random between sets of nearest neighbors anywhere on the
lattice.  We thus arrive at about \( 2000 \) sample values per data
point.

\section{Results}

\newcommand{\Rydberg}{{\text{Ry}}}

\subsection{The Clean System}

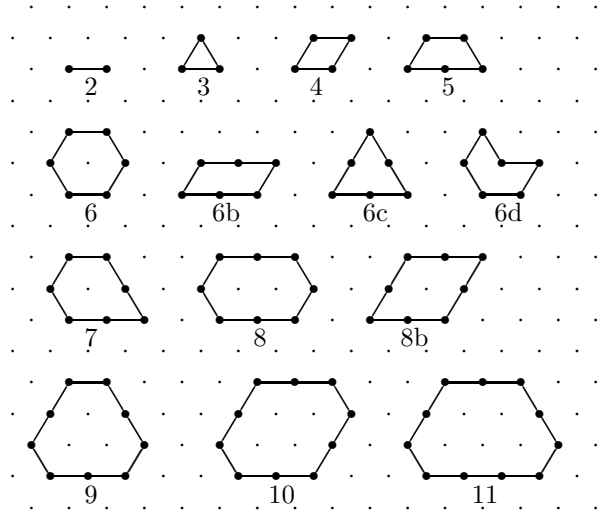
\begin{figure}
\newcommand{\dd}{\circle*{0.05}}
\newcommand{\pp}[1]{\put(#1,0){\circle*{0.2}}}
\newcommand{\lh}[1]{\put(#1,0){\line(1,0){1}}}
\newcommand{\ly}[1]{\put(#1,0){\line(-3,-5){0.5}}}
\newcommand{\lx}[1]{\put(#1,0){\line(3,-5){0.5}}}
\newcommand{\tx}[2]{\put(#1,-0.7){#2}}
\setlength{\unitlength}{0.5cm}
\newsavebox{\onerow}
\savebox{\onerow}(0,0){
        \multiput(1,0)(1,0){16}{\dd}
}
\newsavebox{\tworow}
\savebox{\tworow}(0,0){
        \multiput(1.5,0)(1,0){16}{\dd}
}
\begin{center}
\begin{picture}(18,12)
\put(0,10.8333) {
        \put(0, 2.4999){\usebox{\tworow}}
        \put(0, 1.6666){\usebox{\onerow}
                \ly{6}\lx{6} \pp{6}                        
                \lh{9}\ly{9}\ly{10} \pp{9}\pp{10}          
                \ly{12}\lx{13}\lh{12} \pp{12}\pp{13}       
        }
        \put(0,  .8333){\usebox{\tworow}
                \pp{2.5}\pp{3.5}\lh{2.5}                      
                \pp{5.5}\pp{6.5}\lh{5.5}                      
                \pp{8.5}\pp{9.5}\lh{8.5}                      
                \pp{11.5}\pp{12.5}\pp{13.5}\lh{11.5}\lh{12.5} 
                \tx{2.9}{2}\tx{5.9}{3}\tx{8.9}{4}\tx{12.4}{5}
        }
        \put(0, 0     ){\usebox{\onerow}}
}

\put(0, 7.4999) {
        \put(0, 2.4999){\usebox{\tworow}
                \lh{2.5}\ly{2.5}\lx{3.5} \pp{2.5}\pp{3.5}      

                \ly{10.5}\lx{10.5} \pp{10.5}                   
                \ly{13.5}\lx{13.5} \pp{13.5}                   
        }
        \put(0, 1.6666){\usebox{\onerow}
                \lx{2}\ly{4} \pp{2}\pp{4}                      
                \ly{6}\ly{8}\lh{6}\lh{7} \pp{6}\pp{7}\pp{8}    
                \ly{10}\lx{11} \pp{10}\pp{11}                  
                \lx{13}\ly{15}\lh{14} \pp{13}\pp{14}\pp{15}    
        }
        \put(0,  .8333){\usebox{\tworow}
                \lh{2.5} \pp{2.5}\pp{3.5}                      
                \lh{5.5}\lh{6.5} \pp{5.5}\pp{6.5}\pp{7.5}      
                \lh{9.5}\lh{10.5} \pp{9.5}\pp{10.5}\pp{11.5}   
                \lh{13.5} \pp{13.5}\pp{14.5}                   
                \tx{2.9}{6}\tx{6.3}{6b}\tx{10.3}{6c}\tx{13.8}{6d}
        }
        \put(0,      0){\usebox{\onerow}}
}

\put(0, 4.1666) {
        \put(0, 2.4999){\usebox{\tworow}
                \lh{2.5}\ly{2.5}\lx{3.5} \pp{2.5}\pp{3.5}                        
                \lh{6.5}\lh{7.5}\ly{6.5}\lx{8.5} \pp{6.5}\pp{7.5}\pp{8.5}        
                \ly{11.5}\ly{13.5}\lh{11.5}\lh{12.5} \pp{11.5}\pp{12.5}\pp{13.5} 
        }
        \put(0, 1.6666){\usebox{\onerow}
                \lx{2}\lx{4} \pp{2}\pp{4}                      
                \lx{6}\ly{9} \pp{6}\pp{9}                      
                \ly{11}\ly{13} \pp{11}\pp{13}                  
        }
        \put(0,  .8333){\usebox{\tworow}
                \lh{2.5}\lh{3.5} \pp{2.5}\pp{3.5}\pp{4.5}        
                \lh{6.5}\lh{7.5} \pp{6.5}\pp{7.5}\pp{8.5}        
                \lh{10.5}\lh{11.5} \pp{10.5}\pp{11.5}\pp{12.5}   
                \tx{2.9}{7}\tx{7.4}{8}\tx{11.3}{8b}
        }
        \put(0,      0){\usebox{\onerow}}
}

\put(0, 0     ) {
        \put(0, 3.3333){\usebox{\tworow}
                \lh{2.5}\ly{2.5}\lx{3.5} \pp{2.5}\pp{3.5}     
                \lh{7.5}\ly{7.5}\lx{9.5}\lh{8.5} \pp{7.5}\pp{8.5}\pp{9.5} 
                \lh{12.5}\ly{12.5}\lx{14.5}\lh{13.5} \pp{12.5}\pp{13.5}\pp{14.5} 
        }
        \put(0, 2.4999){\usebox{\onerow}
                \ly{2}\lx{4} \pp{2}\pp{4}                     
                \ly{7}\ly{10} \pp{7}\pp{10}                   
                \ly{12}\lx{15} \pp{12}\pp{15}                 
        }
        \put(0, 1.6666){\usebox{\tworow}
                \lx{1.5}\ly{4.5} \pp{1.5}\pp{4.5}             
                \lx{6.5}\ly{9.5} \pp{6.5}\pp{9.5}             
                \lx{11.5}\ly{15.5} \pp{11.5}\pp{15.5}         
        }
        \put(0, 0.8333){\usebox{\onerow}
                \lh{2}\lh{3} \pp{2}\pp{3}\pp{4}               
                \lh{7}\lh{8} \pp{7}\pp{8}\pp{9}               
                \lh{12}\lh{13}\lh{14} \pp{12}\pp{13}\pp{14}\pp{15}   
                \tx{2.9}{9}\tx{7.8}{10}\tx{13.2}{11}
        }
        \put(0, 0     ){\usebox{\tworow}}
}
\end{picture}
\end{center}
\caption{The most important exchange paths (and some less important
  ones). The paths for \( n=12,14 \) and \( 16 \) can be found by
  adding a ring of particles around the \( n=6,8 \) and \( 12 \)
  diagrams, in the same way as the \( n=8,9,10 \) and \( 11 \)
  diagrams can be derived from \( n=2,3,4 \) and \( 6 \).}
\label{fig: exchanges}
\end{figure}

Here we present results for a large number of exchange processes in the
absence of disorder, including all those that are relevant at low
densities. The exchange paths are shown schematically in Fig.~\ref{fig: exchanges},
and the corresponding values of the dimensionless action \(
\tilde{S}_{n} \) are listed in Table~\ref{tab: exchanges}.
Roughly speaking, the action depends both on the number of particles
involved, and on the smoothness of the exchange paths. Kinks in the path
are penalized, since they lead to intermediate configurations with high
potential energy. This is also the reason for the relatively high value
of $\tilde S_2$. For the smoothest exchange paths with \( n\ge 8 \)
the action increases roughly linear with \( n \) (see Fig.~\ref{fig:
action}). We have, approximately,

\begin{equation}
\tilde{S}_{n}\simeq 0.44+0.22n\qquad (n\ge 8).
\end{equation}

We did not consider processes where a particle tunnels to a location
other than a nearest-neighbor site, since the action for such
processes will be considerably higher. The exchange frequency depends
exponentially on the action, so that even a relatively small increase in
\( \tilde{S} \) can suppress \( J \) quite substantially.

\begin{table}
\begin{tabular}
{ c c | c c | c c | c c }
$n$ & $\tilde S_n$ & $n$ & $\tilde S_n$ &
$n$ & $\tilde S_n$ & $n$ & $\tilde S_n$ \\
\hline
2 & 1.644 & 6b & 2.134 & 8b & 2.764 & 14 & 3.514 \\
3 & 1.526 & 6c & 2.526 & 9  & 2.410 & 16 & 3.934 \\
4 & 1.662 & 6d & 2.294 & 10 & 2.623 &&\\
5 & 1.911 & 7  & 2.220 & 11 & 2.862 &&\\
6 & 1.783 & 8  & 2.188 & 12 & 3.095 &&\\
\end{tabular}
\caption{The dimensionless action \( \tilde S_{n} \)
  for various exchange processes, see Fig.~\ref{fig: exchanges}.}
\label{tab: exchanges}
\end{table}

\begin{figure}
  \centerline{\epsfxsize=3.4in \epsffile{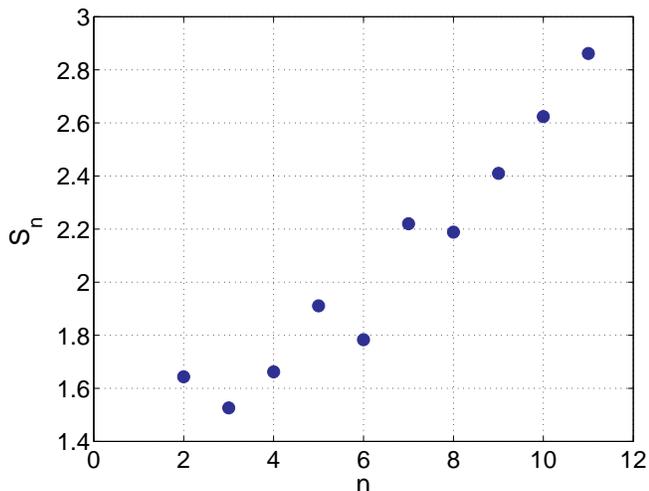}}
\caption{The classical action for the first ten ring exchange
processes.} \label{fig: action}
\end{figure}

\begin{figure}
  \centerline{\epsfxsize=3.4in \epsffile{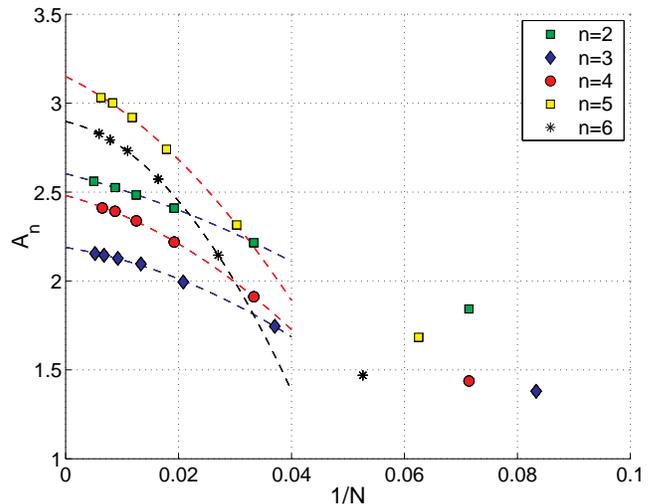}}
\caption{Scaling of the prefactor for two- to six-particle ring
exchanges with inverse system
size. $N$ is the number of particles that are allowed to move.}
\label{fig_prefscaling} \end{figure}

We define the dimensionless prefactor $\tilde A_n$ by writing
\begin{equation}
    \frac{J_n}{\Rydberg} = \tilde A_n \, r_{s}^{-5/4} \,
  \sqrt{\frac{\tilde{S}_n}{2\pi }} \, e^{-r_{s}^{1/2}\tilde{S}_n},
\end{equation}
where \( \Rydberg=e^{2}/2\epsilon a_{B} \) is the effective Rydberg
constant.
In contrast to the classical action, the prefactor shows a strong
dependence on the system size, as can be seen
in Fig.~\ref{fig_prefscaling}. The $N$-dependence fits well to a
scaling form
\begin{equation}
    \tilde A_n(N) = a_\infty - \frac{a_1}{N} - \frac{a_2}{N^2},
\end{equation}
from which we can extract the values for the infinite system:
$\tilde A_2 = 2.60$, \cite{noteA2}
$\tilde A_3 = 2.19$,
$\tilde A_4 = 2.48$,
$\tilde A_5 = 3.15$, and
$\tilde A_6 = 2.90$.

\subsection{Results in the Presence of Disorder}

Although the technique outlined in Sec.~\ref{sec_num_prefactor}
allows us in principle to calculate the prefactor in the presence of
impurities as well, we maintain the viewpoint that all qualitatively
important changes in the exchange frequencies will be caused by
variations of the classical action with disorder, and that the prefactor
$A_n$ depends only weakly on disorder. This is by no means guaranteed,
and in particular the dependence of the typical phonon frequencies on
the disorder has to be investigated further. Note also that close to
the melting transition anharmonicities that are not accounted for in
the instanton approximation may soften the phonon modes significantly.
We assume, however, that the disorder dependence of the exchange
frequencies is dominated by the exponential factor. A small change in
either the action or the prefactor causes a relative change
\begin{equation}
  \frac{\delta J_n}{J_n} = \frac{\delta \tilde A_n}{\tilde A_n}
   + \left( \frac{1}{2} - r_s^{1/2} \tilde{S}_n \right)
     \frac{\delta\tilde{S}_n}{\tilde{S}_n}
\end{equation}
in the exchange frequency. For
moderately large  values of \( r_{s} \) the last term will give
the dominant contribution.


\newcommand{\D}{d/a}

\begin{figure}
  \centerline{\epsfxsize=3.4in \epsffile{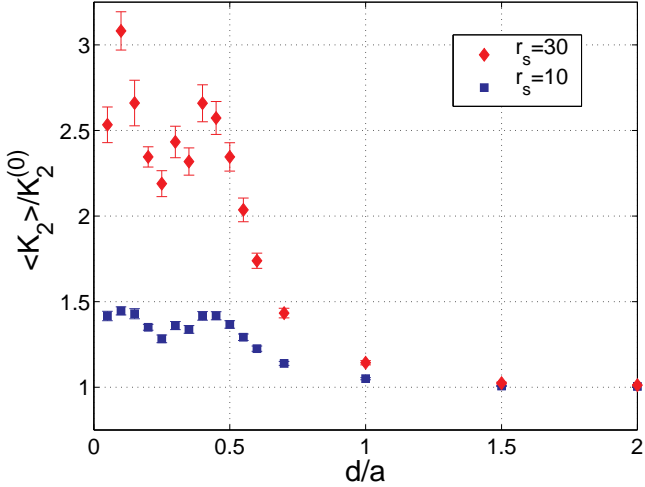}}
\caption{Two-particle exchange frequency relative to its value for the
  clean system, as a function of impurity layer distance}
\label{fig: K2}
\end{figure}

To be explicit, we define the "reduced" exchange constants
\begin{equation}
  \label{eq_defK}
  K_{n}:=\sqrt{\tilde{S}_{n}} \; e^{-r_{s}^{1/2}\tilde{S}_{n}}
\end{equation}
and study their dependence on disorder. Figs.~\ref{fig: K2}--\ref{fig:
K4} show the disorder averages of $K_n$ for $n=2$, $3$ and $4$,
normalized by their values $K_n^{(0)}$ for the clean system. Here $d/a$
is the distance to the impurity layer in units of the lattice constant
of the clean Wigner crystal. The impurity concentration is taken to be
\( x=1/8 \) impurity ions per electron. The system size used for
determining the equilibrium configuration is \( N=48 \).

\begin{figure}
  \centerline{\epsfxsize=3.4in \epsffile{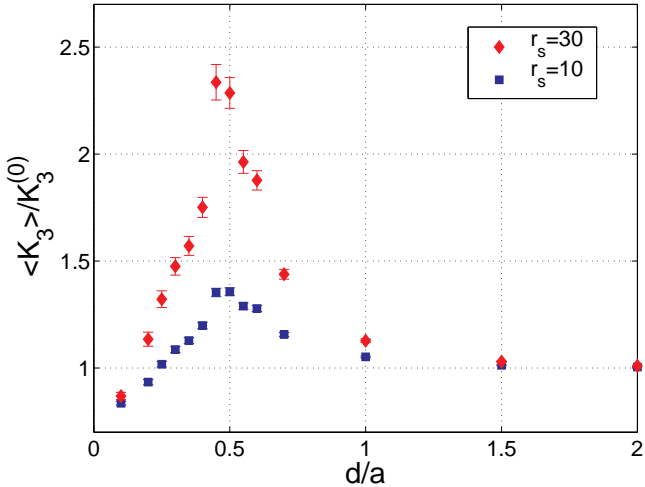}}
\caption{Three-particle exchange frequency relative to its value for
  the clean system, as a function of impurity layer distance}
\label{fig: K3}
\end{figure}

While the impurities are practically of no effect for \( \D \gtrsim 1 \),
in each case we see an enhancement of the average exchange frequency by
up to a factor of $3$ (at $r_s = 30$; at lower carrier densities the
enhancement will be considerably larger) at \( \D\simeq 0.5 \).
For smaller values of \( \D \) three- and four-particle exchange
frequencies decline again, and for \( \D\to 0 \) $K_3$ even falls
below its original value. In the following we
give an interpretation of these results; more details on the frequency
distribution can be found in Appendix~\ref{sec_distributions}.

In Fig.~\ref{fig: rms_displacement} we show the deviation
of the electrons' equilibrium configuration from the ordered lattice,
due to disorder. The average (static) root mean
square displacement shows a sharp increase around the
same value \( \D\simeq 0.5 \) at which the exchange frequencies peak.
We claim that both signatures are due to a structural crossover that
will be investigated in more detail in the following section. In the
crossover region fluctuation effects are amplified, which results in a
softening of the potential barriers and an increase in the variance of
the action of a instanton process. The increase in variance leads to an
increase of the average exchange frequency, due to the positive
curvature of \( J_n(\tilde{S}) \).

\begin{figure}
  \centerline{\epsfxsize=3.4in \epsffile{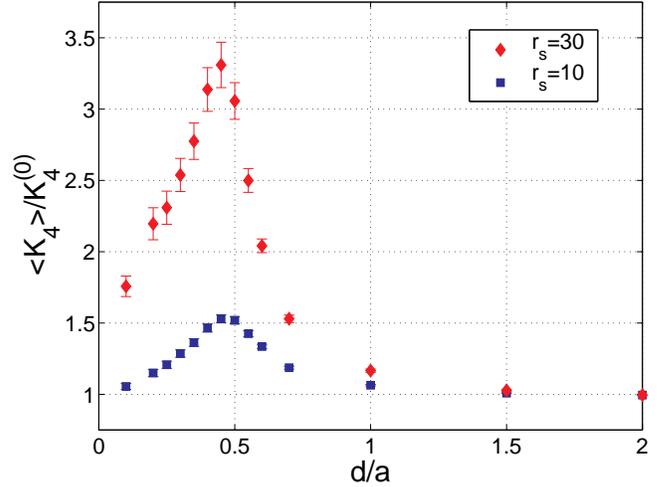}}
\caption{Four-particle exchange frequency relative to its value for
  the clean system, as a function of impurity layer distance}
\label{fig: K4}
\end{figure}

While three-and four-particle exchange frequencies fall off as we
decrease $\D$ below $0.5$, $K_2$ remains enhanced by a factor of about
$2.5$ with respect to the clean system. Hence, if the distance to the
impurity layer becomes very small, the two-particle exchange will
dominate the magnetic properties and enhance antiferromagnetic
correlations. Of course antiferromagnetic order can be
realized only locally. Nevertheless, this raises the
possibility of a magnetic crossover signature that can, in principle,
be picked up experimentally.

Presumably this enhancement of the two-particle exchange
frequency is due to impurities mediating spin singlet
correlations between the electrons of the Wigner glass (see also the
next section). If an electron is trapped by an impurity charge, its
repulsive interaction with neighboring electrons will be greatly reduced
by the impurity potential. Therefore exchange paths in which a second
electron moves very close will not be suppressed as much in the
partition function. Unless the impurity concentration is very high ($x
\simeq 1$), this mechanism will not apply to $n>2$ exchange processes.

\begin{figure}
  \centerline{\epsfxsize=3.4in \epsffile{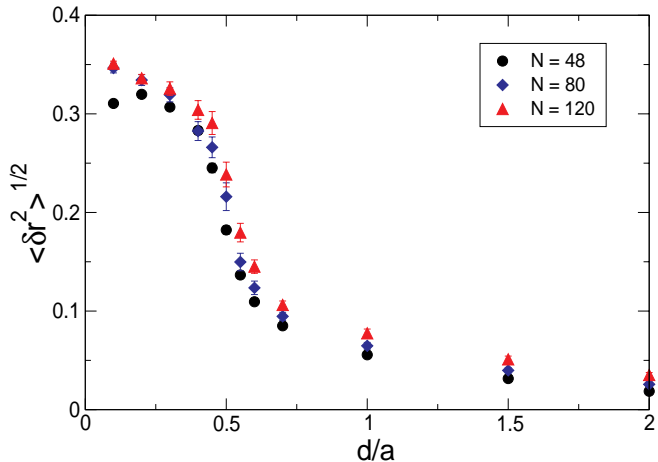}}
\caption{Root mean square displacement of the electrons
  from their equilibrium positions, due to disorder. The three curves
  are for different system sizes.}
\label{fig: rms_displacement}
\end{figure}

\section{The Structural Crossover}

For large \( \D \) the influence of the impurities is weak, and on a
local scale the lattice will be only slightly distorted (Fig.~\ref{fig:
snapshots1}). In the opposite limit ($\D\to 0$), some electrons
will be trapped in the potential wells created by impurity charges.
These electrons are effectively removed from the lattice. The
electron-impurity pairs now appear as dipoles with a dipole
strength proportional to \( d \), and therefore the effective disorder
strength decreases as \( d \) becomes smaller.  The remaining electrons
will rearrange themselves to form a \emph{local} Wigner lattice with an
electron density less than that of the clean system  (Fig.~\ref{fig:
snapshots2}). In the classical limit, and for \( d \)
exactly equal to zero, the remaining
electrons again form a perfect Wigner crystal, but with its electron
density reduced by a factor \( 1-x \), where \( x \) is the impurity
concentration.

\begin{figure}
  \centerline{\epsfxsize=3.4in \epsffile{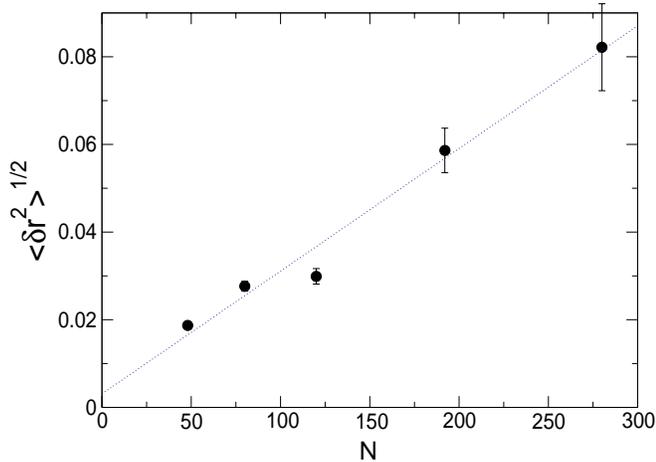}}
\caption{rms deviation of electron coordinates from their equilibrium
  positions, due to disorder, at \protect\( \D=2\protect \), for
  various system sizes. The dotted line is a least-square fit.}
\label{fig: scaling}
\end{figure}

The two limits \( \D=\infty \) and \( \D=0 \) therefore correspond to
two distinct structural phases of the system, with different
translational symmetries. However, no phase transition involving
these symmetries can occur at any finite value of \( \D \). It
has been known for a long time that disorder destroys any spatial
long-range order in two-dimensional systems, \cite{Larkin} and more
recently it has been shown that quasi-long range order, such as in a
Bragg glass, cannot survive either. \cite{DSFisher} To confirm the
absence of long range order we show in Fig.~\ref{fig: scaling} the
dependence of the rms deviations on the system size for weak disorder.
The deviations increase linearly with the particle number, hence
quadratically in the linear dimensions.

Locally, however, we can observe a sharp crossover between the two
structural ``phases''. The correlation length will be strongly
enhanced in the crossover region, but will remain finite due to a long
distance cutoff imposed by disorder.  This enhancement of the
correlation length will give rise to similar effects as are usually
connected with phase transitions, such as the softening of phonon modes
and enhancement of fluctuations. Of course these effects can only
be observed locally.

\begin{figure}
  \centerline{\epsfxsize=3.4in \epsffile{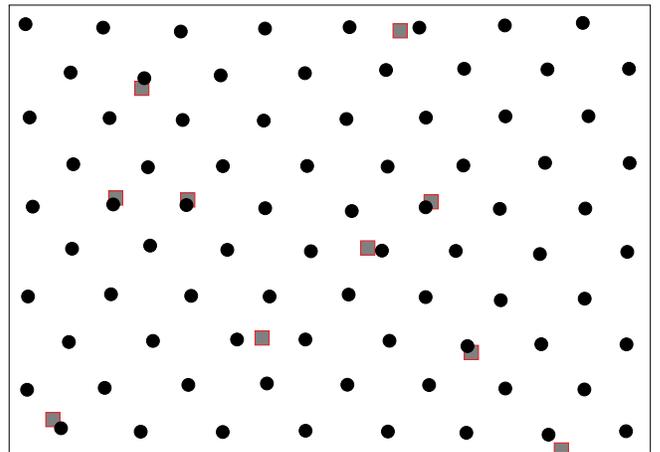}}
  \caption{Snapshot of the classical equilibrium electron configuration
  at \( \D=0.7 \). The filled
  circles show the electron positions, while the checked squares
  indicate the locations of the impurities.}
\label{fig: snapshots1}
\end{figure}

\begin{figure}
\centerline{\epsfxsize=3.4in \epsffile{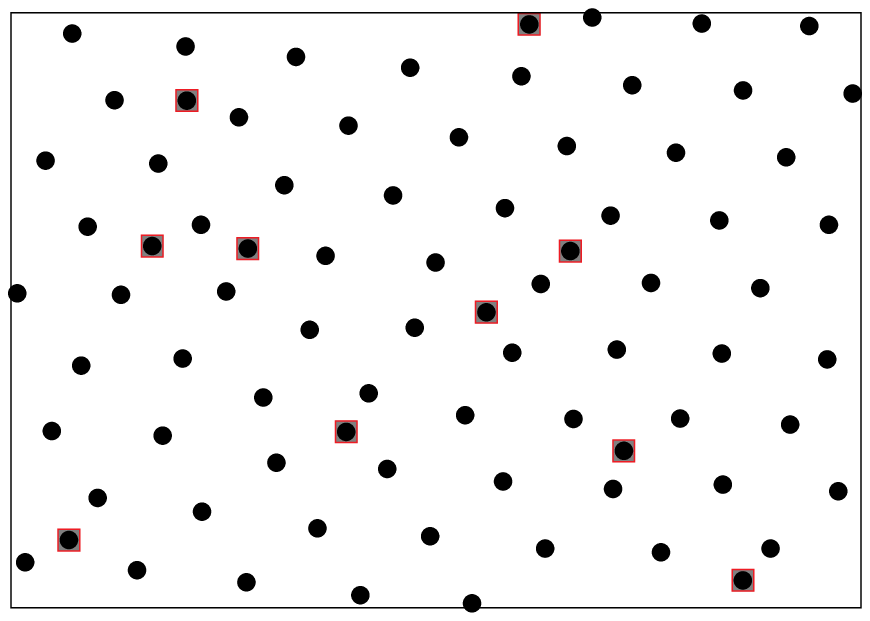}}
\caption{Snapshot of the classical equilibrium electron configuration
  at  \( \D=0.1 \). The filled
  circles show the electron positions, while the checked squares
  indicate the locations of the impurities.}
\label{fig: snapshots2}
\end{figure}

\section{Conclusions}

\begin{figure}
\centerline{\epsfxsize=3.2in \epsffile{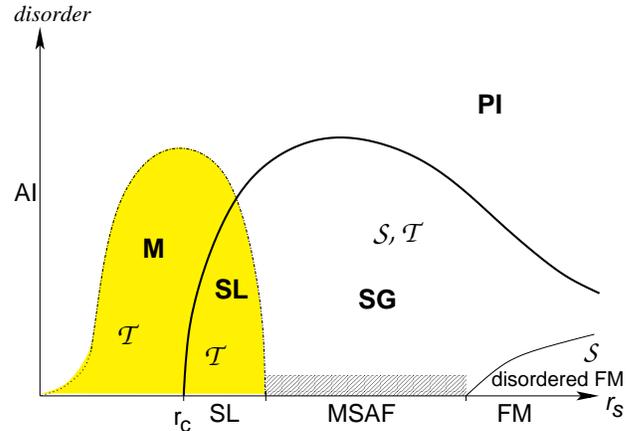}}
\caption{Speculative phase diagram in the $r_s$ - disorder plane. The
various phases are: Anderson insulator (AI), paramagnetic
(Efros-Shklovskii) insulator (PI), spin glass (SG), spin liquid (SL),
various multi-sublattice antiferromagnetic states (MSAF), ferromagnet
(FM), and a metallic state with broken time-reversal symmetry (M).
The diagonally shaded area is a crossover region with intermediate-range
MSAF correlations. $\cal S$ and $\cal T$ indicate broken spin-rotational
and time-reversal symmetry, respectively. Not included is a possible
superconducting state with broken $U(1)$ symmetry.}
\label{fig_phasediagram} \end{figure}

We have attempted to characterize the effective low energy spin Hamiltonian for a
disordered Wigner crystal, or a Wigner glass and have shown that disorder can make a
qualitative difference. In particular,  disorder
can cause an enhancement of the two-particle exchange frequency relative to the other
exchange frequencies, thereby making a possible ferromagnetically ordered state less
likely and a spin liquid phase more likely. The solution of such a complex, competing multiparticle
spin Hamiltonian including disorder is a formidable many body problem. Nonetheless, it would be
surprising if this Hamiltonian did not exhibit a multiplicity of competing phases in the ground
state. We now present a speculative phase diagram   in the
$r_s$ vs.~disorder plane (see Fig.~\ref{fig_phasediagram}), based primarily on symmetry
considerations that can provide some guidance in the future.

Let us first focus on the zero disorder axis: In the limit of very low
densites ($r_s \to \infty$), three-particle exchange will be most
relevant, leading to a state with ferromagnetic long-range order (FM).\cite{Herring}
Upon increasing the density, two- and four-particle exchange will
frustrate the ferromagnet, until ferromagnetic order disappears at a
critical density $r_F$. Within a mean-field approximation, the
(truncated) effective spin Hamiltonian can exhibit a variety of
multi-sublattice antiferromagnetic phases (MSAF).\cite{Misguich} There is also the
possibility of a spin liquid phase (SL).\cite{Misguich1} For even higher densities, the
Wigner crystal will quantum melt at $r_s = r_c$.

While the ferromagnet with disorder averaged order parameter $\overline{<\Svec_i> }\ne 0$ can
survive in the presence of weak disorder, the various MSAF states are tied to the existence of
translational symmetry. As noted earlier, crystalline order will be destroyed on long length scales
by any amount of disorder in two dimensions. Hence, no multi-sublattice antiferromagnetic phases can
exist in the disordered system. Similarly, various dimerized broken symmetry states cannot be
distinct states of matter either.\cite{Sachdev}  We expect a crossover region, with short-range
antiferromagnetic correlations, to a spin glass phase (SG) at $T=0$, although there may not be a finite
temperature spin glass transition in $d=2$. (In fact, numerical finite-size diagonalizations in  
simpler randomly frustrated spin-1/2 nearest neighbor Heisenberg models have been argued to exhibit spin
glass behavior in the ground state.\cite{Oitmaa}) The complexity of this problem can be visualized by the
fact that quenched disorder in such a frustrated multiparticle exchange Hamiltonain  at
$T=0$ appears as infinitely long ranged correlated disorder in the imaginary time direction in the field
theoretical description. In any case, the spin glass phase should be characterized by a nonvanishing
Edwards-Anderson order parameter, that is, the disorder average
$\overline{ <\Svec_i>^2 }$
is nonzero, but $\overline{ <\Svec_i> }=0$.

It is also interesting to construct a chiral order parameter. Let $\Rvec_1$, $\Rvec_2$, and $\Rvec_3$ be
the vertices of a triangular plaquette and define $\Phi(\Rvec)=\langle
\Svec(\Rvec_1)\cdot\Svec(\Rvec_2)\times\Svec(\Rvec_3)\rangle$, where $\Rvec$ is  the center of the
triangle.  In the spin glass phase,
$\overline{\Phi({\bf R})}=0$, but parasitically  $\overline{\Phi^2({\bf R})}\ne 0$. This suggests a
transition to an  adjacent phase in which the Edwards-Anderson order parameter is zero, but
$\overline{\Phi^2({\bf R})}\ne 0$. This  is a new state of matter
with broken time reversal symmetry, which we shall label to be the random flux state.
It is interesting
to ask what a prototype model could be.  It is tempting to
speculate that this is a variant of the random flux model.
Although there is some evidence  of  a metal-insulator transition in the random
flux model, separating a
$\cal T$-broken metallic state from a $\cal T$-broken insulating state, the
evidence to the contrary also exists.\cite{RandomFlux} The resolution of this
controversy should be an important advance.

Next, we look at the strong-disorder, low-density region. Since the
exchange frequencies decrease exponentially with $r_s^{1/2}$, the
characteristic energy scale of disorder will be the largest energy scale, so that the
carriers are independently trapped in some minima of the disorder
potential. The resulting state will be a paramagnetic
(Efros-Shklovskii) insulator. Since no symmetries are broken in
this state, it can be continuously connected to the noninteracting
disordered electron system, which is an Anderson insulator.

A superconducting state with broken $U(1)$ symmetry \cite{Phillips} is
possible in principle. It is not clear to us, however, where in the phase diagram such a
superconducting state should occur, therefore it is not included in the phase
diagram shown in Fig.~(\ref{fig_phasediagram}). We certainly do not imply that such a
phase is impossible.

In none of the  phases 
involving broken
$\cal T$ and
$\cal S$,  the impurity potential can  couple to the order parameter as a ``random field". Rather,
the effect of the potential scattering due to impurities is to randomize the exchange constants.
Similar to rigorous results known in classical statistical mechanics,\cite{Wortis} we can argue that
these broken symmetry transitions in the ground state are necessarily continuous.\cite{Berker} 
Thus, scaling
must hold  at these quantum phase transitions,  and the signature of quantum criticality should be
observable at finite, but low temperatures. In contrast, the Wigner crystal transition
of a pure system is a first order transition at which scaling could not possibly hold.

\acknowledgments

This work was supported by NSF-DMR-9971138. We would like to thank
H. K. Nguyen, S. Kivelson and  C. Nayak for discussions. More recently, C. de Chamon
has indicated to us the possibilty of competing ferromagnetic and p-wave
superconducting states. We would like to thank him for interesting discussions and
correspondence. S. C. would also like to thank the Aspen Center for Physics where 
a part of the work was carried out.

\appendix

\section{Accuracy Issues}
\label{sec_accuracy}

In Table~\ref{table_timeslices} we see how the classical action depends
on the number of time slices $\nslices$ used in the discretized form
(\ref{eqn_discreteaction}). We use the
three-particle exchange, with 27 mobile particles, as an example. The
errors for other exchange processes are comparable. Extrapolating to
$M=\infty$ yields $\tilde S_3 \simeq 1.5848$. Errors
are given with respect to this value.

\begin{table}
\begin{tabular}{ccc}
  $\nslices$ &   $\tilde S_3$    & err    \\
  \hline
   2 & 1.1924 & -25\%  \\
   4 & 1.4919 & -5.9\% \\
   6 & 1.5455 & -2.5\% \\
   8 & 1.5638 & -1.3\% \\
  12 & 1.5763 & -.54\% \\
  16 & 1.5804 & -.28\% \\
  32 & 1.5840 & -.05\% \\
\end{tabular}
\caption{Dependence of the classical action on the number of time
  slices used in the calculation.}
\label{table_timeslices}
\end{table}

Table~\ref{table_mobile} shows the dependence on the number of particles
allowed to move in the exchange. Here $L$ is the number of layers added
around the interchanging particles. The number of time
slices is $\nslices = 12$. Extrapolation to $L=\infty$ yields $\tilde
S_3 \simeq 1.5558$. Since the errors introduced by finite $\nslices$ and
finite $\nmobile$ are of opposite sign, the choices $\nslices=16$ and
$\nmobile\simeq 80$ for the clean system should yield
accurate results to within $\sim 0.2$ percent. For the disordered
system, where we used $\nslices=8$ and $\nmobile\simeq 32$, we expect
the systematic errors to be on the order of $1$ percent.

\begin{table}
\begin{tabular}{cccc}
  $\nmobile$ & $L$ &   $\tilde S_3$   & err  \\
  \hline
   3 & 0 & 1.93334 &  24\% \\
  12 & 1 & 1.65910 & 6.6\% \\
  27 & 2 & 1.57627 & 1.3\% \\
  46 & 3 & 1.56166 & .38\% \\
  69 & 4 & 1.55762 & .11\% \\
  96 & 5 & 1.55628 & .03\% \\
\end{tabular}
\caption{Dependence of the classical action on the number of electrons
  allowed to move.}
\label{table_mobile}
\end{table}

\section{Prefactor in One Dimension}
\label{sec_doublewell}

Here we apply the method introduced in Sec.~\ref{sec_num_prefactor}
for numerical calculation of the prefactor to tunneling in a double
well potential in one dimension, which can be solved exactly. This
serves as a useful test for the validity of the technique. The double
well potential is given by \begin{equation}   V(x) = (x^2-1)^2,
\end{equation}
and the mass of the particle is set to $m=1$. According to
Ref.~\onlinecite{Coleman}, a general expression for the prefactor $P$ in
one dimension is
\begin{equation}
\label{eqn_onedimpref}
  P^{-2} = \frac{2}{\Lambda_0}e^{-\omega_0 T_{\tau}},
\end{equation}
where $T_{\tau}$ is the length of the time slice (we set $T_{\tau}\to\infty$ at
the end of the calculation) and  $\omega_0 = \sqrt{d^2 V/dx^2}|_{x=1}$
is the attempt frequency.  $\Lambda_0$ is the lowest eigenvalue of the
equation
\begin{equation}
  \left[ - \partial_\tau^2 + V(x(\tau)) \right] u(\tau)
  = \Lambda u(\tau),
\end{equation}
which is given by\cite{Coleman}
\begin{equation}
\label{eqn_onedimeval}
  \Lambda_0 = \frac{4\omega_0}{\Sinst} A^2 e^{-\omega_0 T},
\end{equation}
where $\Sinst = \int dx \sqrt{2V(x)}$ is the action along the
classical trajectory, and $A$ is the prefactor for the asymptotic form
of the first time derivative of the classical trajectory:
\begin{equation}
  \dot x_c(\tau) \simeq A e^{-\omega_0|\tau|} \;\;
  \mbox{as} \;\; \tau
\to \pm\infty.
\end{equation}

Integrating the equation of motion (\ref{eqn_of_motion}) yields
\begin{equation}
\label{eqn_onedimmotion}
  x_c(\tau) = \tanh{\sqrt{2}\tau}.
\end{equation}

From Eqs.~(\ref{eqn_onedimpref}), (\ref{eqn_onedimeval}) and
(\ref{eqn_onedimmotion}) we immediately get
\begin{equation}
  P = 4\sqrt{6} \simeq 9.798.
\end{equation}

Table~\ref{table_doublewell} shows results of a numerical computation
using $\nslices$ time slices.
\begin{table}
\begin{tabular}{c|cccccc}
$\nslices$   & $4$ & $8$ & $16$ & $32$ & $64$ & $128$ \\
\hline
$A$ & $10.25$ & $9.970$  & $9.859$ & $9.819$ &
$9.804$ & $9.799$ \\
\end{tabular}
\caption{Prefactor in the double well problem for different numbers of
time slices.}
\label{table_doublewell}
\end{table}
We see that the technique reproduces the
exact result within one percent accuracy for $\nslices \ge 16$.

\section{Distribution of Exchange Frequencies}
\label{sec_distributions}

The dimensionless action $\tilde S_n$ that
enters the expression (\ref{eq_defK}) for the exchange frequencies
depends on the particular realization of disorder, and can therefore be
viewed as a random variable. In most regions of parameter space the
random distribution turns out to be well-described
by a normal distribution
\begin{equation}
  P(\tilde S_n) \simeq \frac{1}{\sqrt{2\pi}\sigma_n}
    e^{ -\frac{(\tilde S_n - \bar S_n)^2}{2\sigma_n^2}}.
\end{equation}

In this approximation, the frequency distribution of $\tilde S_n$, and
thereby of $K_n$, can be reconstructed from two parameters, the mean
action $\bar S_n$ and the standard deviation $\sigma_n$, which are
listed in Table~\ref{table_distribution} for various disorder strengths.

For $\D \lesssim 0.1$ the random distribution acquires a
significant non-gaussian component, hence the corresponding values are
not listed.

\newcommand{\sa}[1]{\bar S_{#1}}
\newcommand{\sd}[1]{\sigma_{#1}}

\begin{table}
\begin{tabular}{c|cc|cc|cc}
$\D$ & $\sa{2}$ & $\sd{2}$ & $\sa{3}$ & $\sd{3}$ & $\sa{4}$ & $\sd{4}$
\\ \hline
2.0 & 1.631 & 0.025 & 1.521 & 0.016 & 1.662 & 0.023 \\
1.5 & 1.633 & 0.043 & 1.519 & 0.031 & 1.659 & 0.042 \\
1.0 & 1.627 & 0.089 & 1.514 & 0.066 & 1.649 & 0.083 \\
0.7 & 1.621 & 0.144 & 1.501 & 0.120 & 1.626 & 0.138 \\
0.6 & 1.614 & 0.171 & 1.490 & 0.153 & 1.600 & 0.177 \\
0.5 & 1.587 & 0.210 & 1.487 & 0.185 & 1.588 & 0.231 \\
0.4 & 1.586 & 0.227 & 1.513 & 0.184 & 1.627 & 0.258 \\
0.3 & 1.572 & 0.234 & 1.533 & 0.189 & 1.678 & 0.290 \\
0.2 & 1.598 & 0.254 & 1.569 & 0.189 & 1.725 & 0.297
\end{tabular}
\caption{Dependence of mean and standard deviation of the
dimensionless action on disorder strength.} \label{table_distribution}
\end{table}


\begin{references}
\bibitem{noteA2} Note that with our definition (\ref{eq_Hpics}) of the
exchange Hamiltonian, $A_2$ contains only the contribution from one of
the two classical paths. The prefactor for the combined
exchange frequency would be $2A_2$.
\bibitem{Berker} A particularly physcially transparent argument given by
  A.~N.~Berker, Physica A \textbf{194}, 72 (1993), is easily
  generalized for a quantum phase transition in the ground state.  All
  that is necessary is to replace the temperature by the coupling
  constant that tunes the phase tarnsition and the free energy by the
  ground state energy.
\end{references}
\end{document}